\documentclass[journal,12]{IEEEtran}
\usepackage[pdftex]{graphicx}
\graphicspath{{../pdf/}{../jpeg/}}
\DeclareGraphicsExtensions{.pdf,.jpeg,.png}
\usepackage[cmex10]{amsmath}
\usepackage{amsthm}
\usepackage{gensymb}
\usepackage{graphicx,color}
\usepackage[dvipsnames]{xcolor}
\usepackage{graphicx}
\usepackage{mathtools}
\usepackage{amssymb,color}
\allowdisplaybreaks
\usepackage{subcaption}
\usepackage{cite}
\usepackage{blindtext}
\usepackage{tcolorbox}
\usepackage{colortbl}

\usepackage{xcolor}
\usepackage{lipsum}
\usepackage{algorithm}
\usepackage{algorithmic}
\usepackage{array} 
\usepackage{makecell}
\usepackage{tabularx}
\usepackage{bm}
\usepackage{siunitx, soul}

\newtheorem{rem}{Remark}
\usepackage{booktabs}
\usepackage{balance}

\usepackage{soul}

\usepackage[shortlabels]{enumitem}

\usepackage[font=scriptsize,labelfont=bf]{caption}

\newcounter{appendx}

\newcommand{\q}[1]{{\bf{#1}}}

\newcommand{\real}[1]{\Re\left({#1}\right)}

\newcommand{\Tr}[1]{\mathrm{Tr}\!\left({#1}\right)}

\renewcommand{\log}[2][]{\mathrm{log}_{#1}\left(#2\right)}

\newcommand{\E}[2][]{\mathbb{E}_{#1}\!\left[{#2}\right]}

\title{Set Transformer-Based Beamforming Design for Cell-Free Integrated Sensing and Communication \vspace{-0mm}}

\author{Ranga Kulathunga, \IEEEmembership{Graduate Student Member, IEEE}, Diluka Galappaththige, \IEEEmembership{Member, IEEE},  Gayan Amarasuriya Aruma Baduge, \IEEEmembership{Senior Member, IEEE} and Chintha Tellambura, \IEEEmembership{Fellow, IEEE}
\thanks{R. Kulathunga and G. A. Aruma Baduge are with the School of Electrical, Computer, and Biomedical Engineering, Southern Illinois University, Carbondale, IL, USA (email: \{ranga.kulathunga, gayan.baduge\}@siu.edu). Their work in part was supported by the U.S. NSF under Grant CCF-2326621.}
\thanks{D. Galappaththige and C. Tellambura are with the Department of Electrical and Computer Engineering, University of Alberta, Edmonton, AB, T6G 1H9, Canada (e-mail: \{diluka.lg, ct4\}@ualberta.ca).
This work, in part, has been submitted to the IEEE 104th Vehicular Technology Conference, 2026, Boston, MA, USA \cite{Kulathunga2026}.
} 
\vspace{-4mm}}

\begin{document}
\bstctlcite{IEEEexample:BSTcontrol}
\maketitle

\begin{abstract}
    Existing cell-free integrated sensing and communication (CF-ISAC) beamforming algorithms predominantly rely on classical optimization techniques, which often entail high computational complexity and limited scalability. Meanwhile, recent learning-based approaches have difficulty capturing the global interactions and long-range dependencies among distributed access points (APs), communication users, and sensing targets. To address these limitations, we propose the first Set Transformer-based CF-ISAC beamforming framework (STCIB). By exploiting attention mechanisms, STCIB explicitly models global relationships among network entities, naturally handles unordered input sets, and preserves permutation invariance across APs, users, and targets. The proposed framework operates in an unsupervised manner, eliminating the need for labeled training data, and supports three design regimes: (i) sensing-centric, (ii) communication-centric, and (iii) joint ISAC optimization. We benchmark STCIB against a convolutional neural network (CNN) baseline and two state-of-the-art optimization algorithms: the convex-concave procedure algorithm (CCPA) and augmented Lagrangian manifold optimization (ALM-MO). Numerical results demonstrate that STCIB consistently outperforms the CNN, achieving substantially higher ISAC performance with only a negligible increase in runtime. For instance, in regime (iii), at $\eta=\num{0.4}$, STCIB improves the sensing and communication sum rates by \qty{14.8}{\percent} and \qty{31.6}{\percent}, respectively, relative to the CNN, while increasing runtime by only \qty{0.26}{\percent}. Compared with CCPA and ALM-MO, STCIB offers significantly lower computational cost while maintaining modest performance gains. In regime (i), for a \qty{3.0}{bps/\Hz} communication threshold, the runtime of STCIB is only \qty{0.1}{\percent} and \qty{0.3}{\percent} of that required by CCPA and ALM-MO, respectively, while improving the sensing sum rate by \qty{4.45}{\percent} and \qty{5.9}{\percent}. Similarly, in regime (ii), for a \qty{0.5}{bps/\Hz} sensing threshold, STCIB requires only \qty{0.1}{\percent} and \qty{0.6}{\percent} of the runtime of CCPA and ALM-MO, while increasing the communication sum rate by \qty{3.2}{\percent} and \qty{3}{\percent}, respectively.
\end{abstract}


\section{Introduction}\label{sec:introduction}
\IEEEPARstart{C}{ell}-free (CF) integrated sensing and communication (ISAC) systems are poised to become a core enabler of next-generation wireless networks \cite{Diluka2025, Diluka2025CFBook, Zhenyao2023, 3GPPISAC2024, Kulathunga2025}. It is driven by the rapid convergence of communication and sensing (C\&S) functions within a unified infrastructure \cite{Diluka2025, Diluka2025CFBook, Kulathunga2025, Zhenyao2023, 3GPPISAC2024}. Moreover, ISAC research has surged recently, revealing that while single-base-station (BS)-centric architectures suffer from inter-cell interference and spatial coverage limitations, the CF concept provides an inherently distributed and cooperative platform for realizing scalable and high-resolution joint C\&S \cite{Kulathunga2025, demirhan2025, mao2024, li2025, zargari2025}. Specifically, the dense deployment of distributed access points (APs), jointly serving users and sensing targets without cell boundaries, allows CF-ISAC to exploit spatial diversity for improved channel hardening, interference mitigation, and multi-static sensing precision \cite{demirhan2025, mao2024, li2025, zargari2025}. Hence, CF-ISAC is expected to play a transformative role in enabling low-latency, high-accuracy, and spectrally efficient operation in emerging applications such as autonomous systems, extended reality, and environment-aware Internet-of-Things networks \cite{Diluka2025, Diluka2025CFBook, Zhenyao2023, 3GPPISAC2024}.

\subsection{Gaps in the CF-ISAC beamforming  literature}

{The ISAC literature is vast and spans multiple research directions. For brevity, only beamforming studies for CF massive multiple-input multiple-output (MIMO) ISAC systems are discussed below.}

Despite the above-mentioned advantages, the performance of CF-ISAC systems critically depends on the design of cooperative AP beamforming, which must simultaneously satisfy user rate requirements, suppress inter-AP interference, and enhance multistatic sensing accuracy \cite{Diluka2025, Diluka2025CFBook, Zhenyao2023}. This leads to a challenging multi-objective optimization problem due to the high-dimensional coupling among distributed APs, users, and sensing targets, as well as the conflicting C\&S objectives. Existing CF-ISAC beamformers can broadly be categorized into optimization-based and learning-based approaches.

\subsubsection{Classical Optimization Approaches}
Semidefinite relaxation (SDR), semidefinite programming (SDP), block coordinate descent (BCD), and augmented Lagrangian methods (ALM) have been widely used for CF-ISAC beamforming design \cite{Kulathunga2025, demirhan2025, mao2024, li2025, rivetti2024, abdelaziz2025, zargari2025}. For instance, \cite{demirhan2025} employs SDR and SDP to maximize sensing signal-to-noise ratio (SNR) under user rate constraints, while \cite{mao2024} addresses sensing-only, communication-only, and joint ISAC designs using Lagrangian duality and fractional transforms combined with BCD and successive convex approximation (SCA). Reference \cite{li2025} minimizes the Cram\'{e}r-Rao lower bound (CRLB) under signal-to-interference-plus-noise ratio (SINR) constraints via second-order cone programming and SDR. In reconfigurable intelligent surfaces (RIS)-assisted CF-ISAC systems, \cite{abdelaziz2025} jointly optimizes RIS phase shifts and transmit beamforming using alternating optimization (AO) with majorization-minimization (MM) and fractional programming (FP), whereas \cite{zargari2025} proposes an ALM-based manifold optimization (MO) framework, termed as ALM-MO, over a complex sphere manifold.

Despite these advances, such approaches typically rely on convex relaxations or surrogate reformulations that may not fully capture the intrinsic nonlinear coupling among APs, users, and sensing targets. Consequently, they require computationally intensive iterative algorithms that are sensitive to initialization and may converge to locally optimal solutions. Furthermore, many formulations assume idealized conditions such as perfect channel state information (CSI), line-of-sight (LoS) propagation, or simplified antenna structures, limiting their applicability to practical large-scale CF networks.

\subsubsection{Learning-Based CF-ISAC Beamforming Approaches}
To improve scalability, \cite{elrashidy2024, demirhan2024, jiang2025, wang2024a, lian2025} have explored learning-based approaches leveraging data-driven models. For example, \cite{elrashidy2024} proposes an unsupervised convolutional neural network (CNN) framework with decentralized neural networks at each AP to maximize sensing SNR and SINR while reducing fronthaul load. However, CNNs have limited capability in modeling long-range interactions among distributed APs \cite{Kocharlakota2026}.

Graph neural networks (GNNs) have therefore been adopted to capture the topology of the CF network. In \cite{demirhan2024}, heterogeneous GNNs are used for joint C\&S optimization, while \cite{jiang2025} introduces dynamic graph learning and a lightweight mirror-based GNN architecture. Similarly, \cite{wang2024a} and \cite{lian2025} employ attention-enhanced GNNs to balance C\&S objectives. Nevertheless, these methods rely on predefined graph connectivity and localized message passing, which restricts information exchange to neighboring nodes and limits the ability to model global AP-user-target interactions.

Consequently, developing a unified CF-ISAC beamforming framework that captures global dependencies while remaining computationally scalable under practical CSI and propagation conditions remains an open challenge.

\subsection{Novelty and Contributions}
{The above discussion reveals two fundamental limitations of the existing solutions. Classical optimization approaches rely on computationally intensive iterative algorithms and often require simplifying assumptions, whereas existing learning-based methods are largely limited to convolutional or graph-based architectures that capture only local interactions. However, CF-ISAC systems inherently involve global dependencies among distributed APs, users, and sensing targets, where network entities form an unordered set and node indexing carries no physical meaning. Modeling such interactions, therefore, requires a framework that operates directly on set-structured data while capturing long-range dependencies in a permutation-invariant manner.}

{This motivates the use of attention-based set learning architectures that enable global information aggregation without imposing artificial structural constraints such as grid layouts or predefined graph connectivity. Conventional Transformers, although attention-based, are designed for ordered sequences and rely on positional encodings that introduce artificial ordering {\cite{lee2019}}. Applying such sequence-based models to CF systems may produce inconsistent outputs under different permutations of physically identical network configurations.}

{To address these limitations, for the first time, this paper introduces a Set Transformer (ST)-driven CF-ISAC beamforming framework (STCIB) that is permutation-invariant and explicitly tailored for unordered set modeling. STCIB leverages the self-attention mechanism, a core component of transformer architectures, in which each element in a set computes attention weights over all other elements, allowing the model to automatically learn which interactions are most relevant for a given task {\cite{vaswani2017}}. The studies in {\cite{jiang2025, wang2024a, lian2025}} apply attention mechanisms only within local neighborhoods defined by predefined graphs in GNNs. In these models, each AP attends only to its directly connected users, restricting interaction modeling to local structures rather than enabling global self-attention across all AP-user-target combinations.}

{In contrast, the proposed STCIB framework employs fully connected global self-attention without relying on predefined topology {\cite{vaswani2017}}. Through this mechanism, each network entity (AP, user, and target) can attend to every other entity, capturing full-order dependencies and enabling comprehensive cooperative modeling across the CF network. This formulation represents a methodological advance rather than an incremental architectural modification, as it introduces a new learning paradigm aligned with the intrinsic set-structured nature of CF-ISAC systems.}

{By replacing iterative optimization with data-driven attention-based inference over unordered sets, STCIB achieves scalable deployment with low online complexity while maintaining near-optimal C\&S performance {\cite{vaswani2017}}. Theoretical validation and simulation results further demonstrate that STCIB achieves superior performance and significantly reduced computational complexity compared to classical optimization and CNN-based approaches.}

\begin{figure}[!t]\centering
    \def\svgwidth{180pt} 
    \fontsize{7}{4}\selectfont 
    \graphicspath{{Figures/}}
    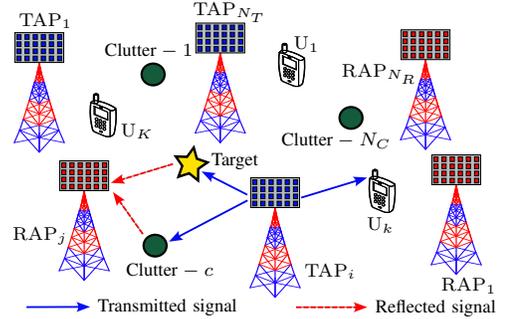 \vspace{-0mm}
    \caption{A system model for CF massive MIMO-assisted ISAC.}
    \label{fig:system_model}\vspace{-3mm}
\end{figure}

To capture a wide range of practical deployments, we consider a generalized CF-ISAC architecture with multi-static sensing, consisting of multiple transmit APs (TAPs) and receive/sensing APs (RAPs) equipped with uniform rectangular planar arrays (URPAs), multiple single-antenna users, and a single target in the presence of multiple clutter sources under spatially correlated Rician fading with channel estimation (Fig.~\ref{fig:system_model}). The goal is to design transmit beamforming strategies under three representative design regimes: sensing-centric (SC), communication-centric (CC), and joint ISAC optimization. The main contributions are summarized as follows:

\begin{enumerate}

\item \textbf{Transformer-driven CF-ISAC beamforming framework.}
We propose the first Set Transformer-based CF-ISAC beamforming framework, termed STCIB. By leveraging attention mechanisms, the proposed architecture captures global interactions among distributed APs, users, and sensing targets while naturally handling unordered network entities and preserving permutation invariance. The framework operates in an unsupervised manner and does not require labeled training data.

\item \textbf{General beamforming formulations for multiple ISAC operating regimes.}
We develop a unified beamforming framework supporting three representative ISAC design regimes: (i) SC design that maximizes the sensing sum rate, $R_s$,  subject to communication QoS and transmit power constraints; (ii) CC design that maximizes the communication sum rate, $R_c$,  subject to sensing requirements and power constraints; and (iii) joint C\&S design that maximizes a weighted sum of $R_s$  and $R_c$ under the total transmit power constraint. These formulations cover a broad spectrum of practical CF-ISAC deployment scenarios.

\item \textbf{Permutation-invariant learning architecture with practical channel estimation.}
To account for realistic network conditions, we incorporate imperfect CSI via a practical channel estimation framework in which uplink channels are locally estimated at each TAP using orthogonal pilots and minimum mean-square error (MMSE) estimation. The STCIB architecture employs Set Transformers, which are designed for unordered set-structured data and enable global AP-user-target interactions via self-attention, making them ideal for CF-ISAC systems.

\item \textbf{Performance gains with substantially reduced computational complexity.}
Extensive simulations demonstrate that STCIB significantly outperforms a CNN-based learning baseline while maintaining nearly identical runtime. For example, in the joint design with $\eta=\num{0.4}$, STCIB improves $R_s$ and $R_c$  by \qty{14.82}{\percent} and \qty{31.60}{\percent}, respectively, with only a \qty{0.26}{\percent} increase in runtime. Compared with optimization-based methods such as the convex-concave procedure algorithm (CCPA) and ALM-MO, STCIB achieves orders-of-magnitude lower runtime while delivering modest performance improvements. For instance, in the SC design in which users are guaranteed a \qty{3.0}{bps/\Hz}, STCIB requires only \qty{0.07}{\percent} and \qty{0.28}{\percent} of the runtime of CCPA and ALM-MO, respectively, while improving $R_s$ by up to \qty{5.91}{\percent}. Furthermore, in the CC design with a sensing threshold of $\qty{0.5}{bps/\Hz}$, the runtime of STCIB lowers runtime by \qty{0.04}{\percent} and \qty{0.63}{\percent} of CCPA and ALM-MO, respectively, while enhancing $R_c$ by \qty{3.19}{\percent} and \qty{2.98}{\percent}.

\item \textbf{Bottom line.} 
These results clearly demonstrate the advantages of the proposed learning-based STCIB scheme, which significantly reduces computational complexity compared to classical optimization methods and achieves superior C\&S performance relative to state-of-the-art learning-based approaches.
\end{enumerate}

This paper goes well beyond our related conference submission \cite{Kulathunga2026}, which considers a CC ISAC beamformer design only. This paper also explores SC and  joint C\&S designs.

\textit{Notation}: Boldface lower case and upper case letters represent vectors and matrices, respectively. For matrix $\mathbf{A}$, $\mathbf{A}^{\rm{H}}$ and $\mathbf{A}^{\rm{T}}$ are Hermitian conjugate transpose and transpose of matrix $\mathbf{A}$, respectively. $\mathbf{I}_M$ denotes the $M$-by-$M$ identity matrix. The Euclidean norm,   absolute value, real part, imaginary part,  Kronecker product, and expectation operators are denoted by $\|\cdot\|$, $|\cdot|$, $\Re(\cdot)$, $\Im(\cdot)$,  $\otimes$, and  $\mathbb{E}\{\cdot\}$,   respectively.   A     complex Gaussian   random vector with mean $\boldsymbol{\mu}$ and covariance matrix $\mathbf{C}$ is denoted by $\sim \mathcal{CN}(\boldsymbol{\mu},\,\mathbf{C})$. Besides, $\mathbb{C}^{M\times N}$ and ${\mathbb{R}^{M \times 1}}$ represent $M\times N$ dimensional complex matrices and $M\times 1$ dimensional real vectors, respectively. Finally, $\mathcal{O}$ expresses the big-O notation.

\section{System, channel and signal models}\label{sec:system_model}
This section outlines the system, channel, and signal models, detailing the communication framework, channel assumptions, mathematical formulations, transmitted and received signals, and essential parameters and variables.

\subsection{System and Channel Models} \label{subsec:system_model}
The CF-ISAC system comprises $N_T$ TAPs and $N_R$ receive/sensing APs to serve $K$ single-antenna communication users and to sense a target (Fig. \ref{fig:system_model}). Each TAP and RAP has $M_T$ and  $M_R$ antennas, respectively. The $i$-th TAP is denoted by $\mathrm{TAP}_i$ for $i\in \{1, \cdots, N_T\}$, the $j$-th RAP by $\mathrm{RAP}_j$ for $j \in \{1, \cdots, N_R\}$, and the $k$-th user by $\mathrm{U}_k$ for $k \in \{1, \cdots, K\}$. Both TAPs and RAPs are equipped with URPAs having $M_{z_1}$ rows and $M_{z_2}$ columns satisfying $M_z = M_{z_1} \times M_{z_2} $, where  $z \in \{T,R\}$. The number of point clutter sources is $N_C.$

\subsubsection{Communication Channel}
The channel between $\mathrm{TAP}_i$ and $\mathrm{U}_k$ is  $\q{h}_{ik} \in \mathbb{C}^{M_T \times 1}$, and modeled as spatially correlated Rician fading to capture the effects of both deterministic LoS and random non-LoS components. It  can be given as
\begin{eqnarray}\label{eqn:Channel_h_ik}
    \q{h}_{ik} = \underbrace{\left( \frac{K_{ik}\varsigma_{ik}}{K_{ik} + 1}\right)^{\!\frac{1}{2}} \bar{\q{h}}_{ik}}_{\q{c}_{ik}} + \underbrace{\left( \frac{\varsigma_{ik} }{K_{ik} + 1}\right)^{\!\frac{1}{2}}}_{e_{ik}^{{1}/{2}}} \tilde{\q{h}}_{ik},
\end{eqnarray}
{where $K_{ik}$ and $\varsigma_{ik}$ are the Rician factor and large-scale fading coefficient of   $\q{h}_{ik}$, respectively. For notational simplicity, we denote $\q{c}_{ik} = (K_{ik}\varsigma_{ik}/(K_{ik} + 1))^{ {1}/{2}} \bar{\q{h}}_{ik} $ and $e_{ik}  = (\varsigma_{ik}/(K_{ik} + 1))$. Moreover, we define $\tilde{\q{h}}_{ik}  \sim\mathcal {CN}_{M_T \times 1}(\boldsymbol{0}_{{M_T} \times 1}, A \tilde{\q{R}}_{ik} ) $, where $\tilde{\q{R}}_{ik} \in \mathbb{C}^{M_T \times M_T}$ is the normalized spatial correlation matrix that can be modeled using the uniform planar array correlation model in {\cite[\textit{Proposition 1}]{Bjornson2020}}, and  $A$ is the area of an antenna element at each TAP.}

In \eqref{eqn:Channel_h_ik},   $\bar{\q{h}}_{ik}$ is the   LoS  component of $\q{h}_{ik}$, and modeled as
\begin{eqnarray}\label{eqn:deterministic channel}
    \bar{\q{h}}_{ik} = \text{vec}\left( \q{b}_{T_1}(\psi_{ik}, \theta_{ik}) \q{b}^{\mathrm{H}}_{T_2}(\psi_{ik}, \theta_{ik})\right), 
\end{eqnarray}
where $\psi_{ik}$ and $\theta_{ik}$ are azimuth and elevation angles of arrival (AoAs) of the channel between $\mathrm{TAP}_i$ and $\mathrm{U}_k$.  The operator $\mathrm{vec}(\cdot)$  vectorizes its argument. Here, $\q{b}_{T_x}(\psi_{ik}, \theta_{ik})$ is defined as \cite{Zhang2014c}
\begin{eqnarray}\label{eqn:b1_and_b2}
    \q{b}_{T_x}(\psi_{ik}, \theta_{ik}) &=&\left[1, \cdots, e^{-j(M_{T_x} - 1) r_x(\psi_{ik}, \theta_{ik}) }\right]^{\mathrm{T}},
\end{eqnarray}
for $x \in \{1,2\}$, and 
\begin{subequations}
    \begin{eqnarray}\label{eqn:r1_and_r2}
        r_1(\psi_{ik}, \theta_{ik}) &=& \pi \cos{(\psi_{ik})} \sin{(\theta_{ik})}, \\ \label{eqn:r1}
        r_2(\psi_{ik}, \theta_{ik}) &=& \pi \sin{(\psi_{ik})} \sin{(\theta_{ik})}.\label{eqn:r2}
    \end{eqnarray}
\end{subequations}

\subsubsection{Sensing Channel} Following the echo signal representation in MIMO radar  \cite{Zhenyao2023, MIMO_radar_book_2008}, the channels between the target and RAPs are assumed to be LoS channels. To this end, the  array response vector for an azimuth angle $\psi_{nm}$ and elevation angle $\theta_{nm}$ can be defined using \eqref{eqn:deterministic channel}-\eqref{eqn:r2} as 
\begin{eqnarray}\label{eqn:array_response_for_sensing}
    \q{a}_{z}(\psi_{nm}, \theta_{nm}) = \text{vec}\left( \q{b}_{z_1}(\psi_{nm}, \theta_{nm}) \q{b}^{\mathrm{H}}_{z_2}(\psi_{nm}, \theta_{nm})\right),
\end{eqnarray}
where $n \in \{i,j\}$, $m \in \{t,c\}$, and $z \in \{T,R\}$. Specifically in (\ref{eqn:array_response_for_sensing}), 	$\q{a}_T(\cdot)$ and $\q{a}_R(\cdot)$ are the array steering vectors of $\mathrm{TAP}_i$ and   $\mathrm{RAP}_j$, respectively. Moreover, $(\psi_{it},\theta_{it})$ and $(\psi_{ic},\theta_{ic})$ are the angles of departure (AoDs) of  $\mathrm{TAP}_i$ for the target and the $c$-th clutter, respectively. Similarly, $(\psi_{jt},\theta_{jt})$ and $(\psi_{jc},\theta_{jc})$  are the AoAs at $\mathrm{RAP}_j$ for the target and the $c$-th clutter, respectively. For notational simplicity, we denote $\q{a}_T(\psi_{im}, \theta_{im})$ and $\q{a}_R(\psi_{jm}, \theta_{jm})$, respectively, as $\q{a}_{T,im}$ and $\q{a}_{R,jm}$, where $m \in \{t,c\}$.

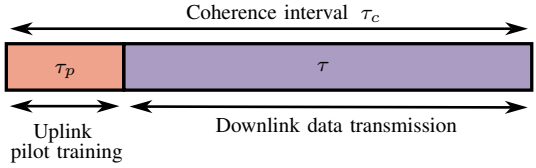
\begin{figure}[!t]\centering
    \def\svgwidth{200pt} 
    \fontsize{8}{4}\selectfont 
    \graphicspath{{Figures/}}
\begingroup%
  \makeatletter%
  \providecommand\color[2][]{%
    \errmessage{(Inkscape) Color is used for the text in Inkscape, but the package 'color.sty' is not loaded}%
    \renewcommand\color[2][]{}%
  }%
  \providecommand\transparent[1]{%
    \errmessage{(Inkscape) Transparency is used (non-zero) for the text in Inkscape, but the package 'transparent.sty' is not loaded}%
    \renewcommand\transparent[1]{}%
  }%
  \providecommand\rotatebox[2]{#2}%
  \newcommand*\fsize{\dimexpr\f@size pt\relax}%
  \newcommand*\lineheight[1]{\fontsize{\fsize}{#1\fsize}\selectfont}%
  \ifx\svgwidth\undefined%
    \setlength{\unitlength}{456.1649737bp}%
    \ifx\svgscale\undefined%
      \relax%
    \else%
      \setlength{\unitlength}{\unitlength * \real{\svgscale}}%
    \fi%
  \else%
    \setlength{\unitlength}{\svgwidth}%
  \fi%
  \global\let\svgwidth\undefined%
  \global\let\svgscale\undefined%
  \makeatother%
  \begin{picture}(1,0.28873981)%
    \lineheight{1}%
    \setlength\tabcolsep{0pt}%
    \put(0,0){\includegraphics[width=\unitlength,page=1]{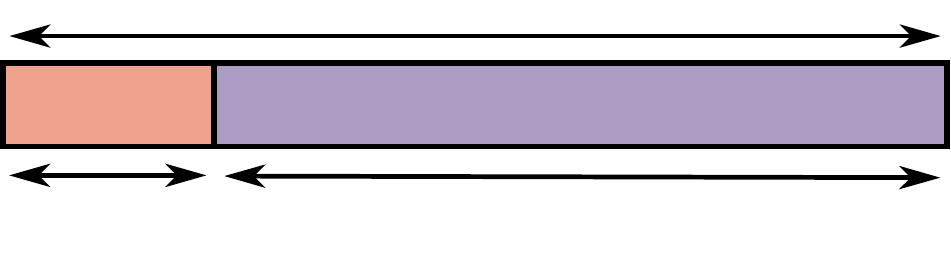}}%
    \put(0.34365241,0.26849633){\color[rgb]{0,0,0}\makebox(0,0)[lt]{\lineheight{1.25}\smash{\begin{tabular}[t]{l}$\text{Coherence interval} \;\; \tau_c$\end{tabular}}}}%
    \put(0.39955249,0.05475769){\color[rgb]{0,0,0}\makebox(0,0)[lt]{\lineheight{1.25}\smash{\begin{tabular}[t]{l}$\text{Downlink data transmission}$\end{tabular}}}}%
    \put(0.05428641,0.04489283){\color[rgb]{0,0,0}\makebox(0,0)[lt]{\lineheight{1.25}\smash{\begin{tabular}[t]{l}$\text{Uplink}$\end{tabular}}}}%
    \put(0.59121195,0.17057946){\color[rgb]{0,0,0}\makebox(0,0)[lt]{\lineheight{1.25}\smash{\begin{tabular}[t]{l}$\tau$\end{tabular}}}}%
    \put(0.09468732,0.16729118){\color[rgb]{0,0,0}\makebox(0,0)[lt]{\lineheight{1.25}\smash{\begin{tabular}[t]{l}$\tau_p$\end{tabular}}}}%
    \put(0.01153872,0.00543338){\color[rgb]{0,0,0}\makebox(0,0)[lt]{\lineheight{1.25}\smash{\begin{tabular}[t]{l}$\text{pilot training}$\end{tabular}}}}%
  \end{picture}%
\endgroup%

    \caption{A transmission time slot.}
    \label{fig_time_slot}\vspace{-3mm}
\end{figure}

\subsection{Uplink Channel Estimation} \label{subsec:Uplink_channel_estimation}
During the uplink pilot phase, all TAPs locally estimate the uplink channels $\q{h}_{ik}$ for $i \in \{1, \cdots, N_T\}$ via pilots sent by $K$ users. {The downlink channels between TAPs and the users are obtained by the channel reciprocity of time-division duplexing (TDD)  mode of operation. TDD enables the system to utilize the same frequency band for uplink and downlink communication, whereas channel reciprocity enables direct acquisition of downlink CSI at the APs from uplink pilot estimation, eliminating the need for additional feedback {\cite{Ngo2017}}.} The $\mathrm{U}_k$'s pilot is denoted by $\boldsymbol{\phi}_k \in \mathbb{C}^{\tau_p}$, where $\tau_p$ is the pilot sequence length within the $\tau_c$ symbols of the coherence interval (Fig.~\ref{fig_time_slot}). It satisfies ${\boldsymbol{\phi}_k} \boldsymbol{\phi}_{k'}^{\mathrm{H}} = 1$ if $k= k'$, and otherwise ${\boldsymbol{\phi}_k} \boldsymbol{\phi}_{k'}^{\mathrm{H}} = 0$ constituting an orthogonal pilot set to avoid deleterious effects of pilot contamination. The pilot signal received at $\mathrm{TAP}_i$ is given by   
\begin{eqnarray}\label{eqn:y_uplink_received_signal}
    \q{Y}_i = \sqrt{\rho_p \tau_p} \sum_{k=1}^{K} \q{h}_{ik} {\boldsymbol{\phi}_k} + \tilde{\q{N}}_i,
\end{eqnarray}
where $\rho_p$ is the uplink pilot transmit SNR, and $\tilde{\q{N}}_i$ $\!\sim\!\mathcal {CN}_{{M_T} \times \tau_p}\left(\q{0}_{{M_T} \times \tau_p}, \q{I}_{{M_T}} \otimes \q{I}_{\tau_p}\right)$ is an additive white Gaussian noise (AWGN) matrix. Upon projecting $\boldsymbol{\phi}_{k}$ onto \eqref{eqn:y_uplink_received_signal}, the post-processed received pilot signal for estimating $\mathrm{U}_k$'s channel at $\mathrm{TAP}_i$  can be given as 
\begin{eqnarray}\label{eqn:y_tilda_kth_user}
    \tilde{\q{y}}_{ik} = \q{Y}_i \boldsymbol{\phi}_k^{\mathrm{H}}  = \sqrt{\rho_p \tau_p} \q{h}_{ik}  + \tilde{\q{n}}_{ik}, 
\end{eqnarray}
where $\tilde{\q{n}}_{ik} = \q{N}_i\boldsymbol\phi^{\mathrm{H}}_k  \sim\mathcal {CN}\left(\q{0}, \q{I}_{M_T}\right)$. The  MMSE estimate of $\q{h}_{ik}$ can be derived as \cite{Kay1993}
\begin{eqnarray}\label{eqn:hik_hat_estimate}
    \hat{\q{h}}_{ik} &=&  \E[]{\q{h}_{ik}} + \q{C}_{h_{ik} \tilde{y}_{ik}} \q{C}^{-1}_{ \tilde{y}_{ik} \tilde{y}_{ik}} (\tilde{\q{y}}_{ik} - \E[]{\tilde{\q{y}}_{ik}}) \nonumber \\ 
    &=& \q{c}_{ik} + \q{C}_{h_{ik} \tilde{y}_{ik}} \q{C}^{-1}_{ \tilde{y}_{ik} \tilde{y}_{ik}} (\tilde{\q{y}}_{ik} - \sqrt{\rho_p \tau_p} \q{c}_{ik}  ),
\end{eqnarray}
where   covariance matrices  $\q{C}_{h_{ik} \tilde{y}_{ik}}$ and $\q{C}_{ \tilde{y}_{ik} \tilde{y}_{ik}}$ are defined as
\begin{eqnarray}\label{eqn:Cov_Cy_tilde}
\!\!\!\!\!\!	\q{C}_{h_{ik} \tilde{y}_{ik}} &=& \sqrt{\rho_p \tau_p} \q{C}_{h_{ik}}   \label{eqn:Cov_Cfak_ytilda} \;\; \text{and}\;\; 
    \q{C}_{ \tilde{y}_{ik} \tilde{y}_{ik}} = \rho_p \tau_p \q{C}_{h_{ik}} + \q{I} ,
\end{eqnarray}
where $\q{C}_{h_{ik}}$ is the covariance matrix of $\q{h}_{ik}$,   and it  can be derived  as 
\begin{eqnarray}\label{eqn:Cov_C_h_ik}
    \q{C}_{h_{ik}} &=& \E{(\q{h}_{ik} - \E{\q{h}_{ik}}) (\q{h}_{ik} - \E{\q{h}_{ik}})^{\mathrm{H}}} \nonumber \\
    &=& \E{e_{ik}^{{1}/{2}} \tilde{\q{h}}_{ik} (e_{ik}^{{1}/{2}} \tilde{\q{h}}_{ik})^{\mathrm{H}}} = e_{ik} \E{\tilde{\q{h}}_{ik} \tilde{\q{h}}_{ik}^{\mathrm{H}}} \nonumber \\
    &=& e_{ik} A \tilde{\q{R}}_{ik}.
\end{eqnarray}
Due to  the orthogonality  criterion of  MMSE estimation \cite{Kay1993}, the true channel $\q{h}_{ik}$ can be written in terms of its MMSE estimate ($\hat{\q{h}}_{ik} $) and estimation  error  $(\boldsymbol{\varepsilon}_{h_{ik}})$ as $	\q{h}_{ik} = \hat{\q{h}}_{ik} + \boldsymbol{\varepsilon}_{h_{ik}}$, where  $\boldsymbol{\varepsilon}_{h_{ik}}$ is  a zero mean Gaussian vector having a  covariance matrix   $\q{C}_{\varepsilon_{h_{ik}}} = \q{C}_{h_{ik}} - \q{C}_{\hat{h}_{ik}}$. Here,  $\boldsymbol{\varepsilon}_{h_{ik}}$ is independent of   $\hat{\q{h}}_{ik}$ \cite{Kay1993}.   The covariance matrix   of $\hat{\q{h}}_{ik}$  can be  derived as
\begin{eqnarray}\label{eqn:Cov_C_h_hat_ik}
    \q{C}_{\hat{h}_{ik}} &=& \E{\Big(\hat{\q{h}}_{ik} - \E{\hat{\q{h}}_{ik}} \Big) \Big(\hat{\q{h}}_{ik} - \E{\hat{\q{h}}_{ik}} \Big)^{\mathrm{H}}} \nonumber \\
    &=& \mathbb{E}[\q{C}_{h_{ik} \tilde{y}_{ik}} \q{C}^{-1}_{ \tilde{y}_{ik} \tilde{y}_{ik}} (\tilde{\q{y}}_{ik} - \sqrt{\rho_p \tau_p} \q{c}_{ik}  )\nonumber \\
    && \times(\q{C}_{h_{ik} \tilde{y}_{ik}} \q{C}^{-1}_{ \tilde{y}_{ik} \tilde{y}_{ik}} (\tilde{\q{y}}_{ik} - \sqrt{\rho_p \tau_p} \q{c}_{ik}  ))^{\mathrm{H}}] \nonumber \\
    &=& \q{C}_{h_{ik}\tilde{y}_{ik}} \q{C}_{\tilde{y}_{ik}\tilde{y}_{ik}}^{-1}\q{C}_{h_{ik}\tilde{y}_{ik}}.
\end{eqnarray}

\subsection{Communication Signal Model} \label{subsec:Signal Model Users}
Downlink data symbols of unit average power are transmitted, i.e., $\q{x} = [x_0, x_1, \cdots, x_K]^{\mathrm{T}} \in \mathbb{C}^{(K+1) \times 1}$, where $x_0$ and $x_k$ are the data symbols assigned for target sensing and $\mathrm{U}_k$, respectively, satisfying $\E{|x_k|^2} = 1$ for $k \in \{0, 1, \cdots, K\}$. The transmitted signal $\q{s}_i \in \mathbb{C}^{M_T \times 1}$ at $\mathrm{TAP}_i$ can be given as
\begin{equation}\label{eqn:tx_signal S_i}
    \q{s}_i  = \sum_{k=0}^{K} \bar{\q{w}}_{ik} x_k = \bar{\q{W}}_i \q{x},
\end{equation}
where $\bar{\q{w}}_{ik} \in \mathbb{C}^{M_T \times 1}$ {is the transmit beamforming vector} at $\mathrm{TAP}_i$, and $\bar{\q{W}}_i = [\bar{\q{w}}_{i0}, \cdots,\bar{\q{w}}_{iK}] \in \mathbb{C}^{M_T \times (K+1)}$. The concatenated received signal at all $K$  users can be written as 
\begin{equation}\label{eqn:combined received signal vector}
    \q{\bar{y}}  = \sum_{i=1}^{N_T}  \q{H}^{\mathrm{H}}_i \q{s}_i + \q{\bar{n}}, 
\end{equation}
where $\q{H}_i = [\q{h}_{i1}, \cdots, \q{h}_{iK}] \in \mathbb{C}^{M_T \times K}$ and $\q{\bar{n}} = [\bar{n}_1, \cdots, \bar{n}_K]^{\mathrm{T}}$ is an AWGN vector. Thereby, the received signal at $\mathrm{U}_k$ can be given by 
\begin{eqnarray}\label{eqn:kth user recieved signal}
&&\!\!\!\!\!\!\!\!\!\!\!\!\!\!\!\!\!\!	\bar{y}_{k}  =    \sum_{i=1}^{N_T}  \q{h}^{\mathrm{H}}_{ik} 	\bar{\q{W}}_{i} \q{x} + \bar{n}_k \nonumber \\
    &=&\!\!\!\!\!\! \sum_{i=1}^{N_T} \! \q{h}^{\mathrm{H}}_{ik} 	\bar{\q{w}}_{ik} x_k +\!\! \sum_{j = 0, j \neq k}^{K}  \sum_{i=1}^{N_T} \! \q{h}^{\mathrm{H}}_{ik} 	\bar{\q{w}}_{ij} x_j \!+\! \bar{n}_k.~\quad
\end{eqnarray}

\noindent By cascading channel vectors into $\q{f}_k = [\q{h}_{1k}^{\mathrm{H}},  \cdots,\q{h}_{N_Tk}^{\mathrm{H}}]^{\mathrm{H}} \in \mathbb{C}^{N_TM_T \times 1}$ and beamforming vectors into $\q{w}_k = [\bar{\q{w}}_{1k}^{\mathrm{H}},  \cdots,\bar{\q{w}}_{N_Tk}^{\mathrm{H}}]^{\mathrm{H}} \in \mathbb{C}^{N_TM_T \times 1}$ for $\mathrm{U}_k$, \eqref{eqn:kth user recieved signal} can be rewritten as
\begin{eqnarray}\label{eqn:kth user recieved signal 2}
    \bar{y}_{k} &=& \q{f}^{\mathrm{H}}_{k} \q{w}_{k} x_k + \sum_{ j = 0, j \neq k}^{K}  \q{f}^{\mathrm{H}}_{k} \q{w}_{j} x_j + \bar{n}_k,
\end{eqnarray}
where the first term is the desired signal component, the second term is the total interference from other users and sensing, and $\bar{n}_k \sim \mathcal{CN}(0,1)$ is an AWGN of $\mathrm{U}_k$. 

\subsection{Sensing Signal Model}\label{sec:radar_sensing_model}
The signals sent by the TAPs are reflected off the target, and the RAPs receive their echoes. They also receive undesired echoes reflected off the clutters. These received echoes are processed by the RAPs to perform functionalities such as detection and localization of the target. The received signal at $\mathrm{RAP}_j$  can be written as 
\begin{eqnarray}\label{eqn:received echo}
    \q{y}_j &=& \sum_{i=1}^{N_T}  \alpha_{ij,t} \q{a}_{R,jt}  \q{a}_{T,it}^{\mathrm{H}}   \bar{\q{W}}_i \q{x} \nonumber \\ 
    &&+ \sum_{c=1}^{N_C}   \sum_{i=1}^{N_T}  \alpha_{ij,c}  \q{a}_{R,jc}  \q{a}_{T,ic} ^{\mathrm{H}} \bar{\q{W}}_i \q{x}  + \q{n}_j,
\end{eqnarray}
where $\alpha_{ij,t}$ and $\alpha_{ij,c}$ are the combined reflection coefficients of the target and the $c$th clutter between $\mathrm{TAP}_i$ and  $\mathrm{RAP}_j$, respectively, and $\q{n}_j$ is an AWGN vector with independent entries distributed as $ \mathcal{CN}(0,1)$. 
{The reflection coefficient $\alpha_{ij,m}$ for $m \in \{t,c\}$ incorporates both propagation attenuation and the radar cross section (RCS) of the corresponding scatterer {\cite{Mark2010RadarBook}}. Specifically, it is modeled as $\alpha_{ij,m} = \sqrt{{\lambda^2 \sigma_{ij,m}^2}\big/({(4\pi)^3 d_{im}^2 d_{jm}^2}}) \varpi_{ij,m}$, where $\lambda$ is the carrier wavelength, $\sigma_{ij,m}^2$ denotes the variance of the RCS corresponding to the propagation path via scatterer $m$, and $d_{im}$ and $d_{jm}$ are the distances from $\mathrm{TAP}_i$ to $m$ and from $m$ to $\mathrm{RAP}_j$, respectively. The normalized RCS component $\varpi_{ij,m}$ captures small-scale fluctuations and is modeled as $\mathcal{CN}(0,1)$.}

By denoting $\bar{\q{G}}_{ij,m} = \alpha_{ij,m} \q{a}_{R,jm}  \q{a}_{T,im}^{\mathrm{H}} \in \mathbb{C}^{M_R \times M_T}$ for $m \in \{t, c\}$, \eqref{eqn:received echo} can be rewritten as
\begin{eqnarray}\label{eqn:received echo 2}
    \q{y}_j   =   \sum_{i=1}^{N_T}  \bar{\q{G}}_{ij,t}    \bar{\q{W}}_i \q{x} + \sum_{c=1}^{N_C}   \sum_{i=1}^{N_T}  \bar{\q{G}}_{ij,c}  \bar{\q{W}}_i \q{x}  + \q{n}_j.
\end{eqnarray}
By denoting $\q{G}_{j,m} = [\bar{\q{G}}_{1j,m}, \cdots, \bar{\q{G}}_{N_Tj,m}] \in \mathbb{C}^{M_R \times N_TM_T}$, where $m \in \{t, c\}$ and $\q{W} = [\q{w}_1, \cdots, \q{w}_{K+1}] \in \mathbb{C}^{N_TM_T \times (K+1)}$, \eqref{eqn:received echo 2} can be rewritten as
\begin{eqnarray}\label{eqn:received echo 3}
    \q{y}_j &=& \q{G}_{j,t}    \q{W} \q{x} + \sum_{c=1}^{N_C}  \q{G}_{j,c}  \q{W} \q{x}  + \q{n}_j,
\end{eqnarray}
where the first term represents the desired target echo component, and the second term is the total clutter interference. {For these sensing response matrices, the target parameters are typically obtained through an initial sensing stage (e.g., beam sweeping or radar probing) that detects dominant reflectors and estimates their angular and propagation parameters {\cite{Zhang2025, Qing2024}}. The estimated parameters are then used to construct $\q{G}_{j,t}$ and $\q{G}_{j,c}$  for subsequent beamforming optimization.}

{Note that while clutter is modeled as a finite set of discrete point scatterers,  distributed or extended scattering regions can occur in practice. Extended clutter can be modeled as a continuum of scatterers distributed over angle, delay, and Doppler,  expressed as $
    \mathbf{y}_{c}= \iiint \alpha(\theta,\tau,f_D)\,\mathbf{a}(\theta)\, s(t-\tau)e^{j2\pi f_D t}\, d\theta\, d\tau\, df_D,$
where $\alpha(\theta,\tau,f_D)$ denotes the distributed clutter reflectivity over the angle-delay-Doppler domain. In practice, this continuous formulation is commonly approximated through sufficiently fine discretization, yielding $
    \mathbf{y}_{c}\approx \sum_{n=1}^{N} \tilde{\alpha}_n  \mathbf{a}(\tilde{\theta}_n)\, s(t-\tilde{\tau}_n)e^{j2\pi \tilde{f}_{D,n} t}, $
where $N$ is selected large enough to accurately capture the spatial extent of clutter  {\cite{Du2023, Pang2024}}. Under this interpretation, extended clutter occupies a structured subspace and can be viewed as a dense collection of virtual scatterers. Therefore, the adopted model in (16) captures the dominant effects of clutter. When extended clutter models are adopted, only the signal-to-clutter-plus-noise ratio (SCNR) used in the sensing-rate metric needs to be modified (Section~{\ref{sec:SCNR}}), while the STCIB  and training remain unchanged. }

\section{Performance  analysis}\label{sec:performace_analysis}
The C\&S performance {is} evaluated by the communication rates of the users and the target's sensing rate at each RAP. 

\subsection{Communication Rate}\label{sec:achievable_rate}
The users use the received signal from all TAPs, i.e., \eqref{eqn:kth user recieved signal 2}, to decode their data. To this end, the downlink achievable rate of $\mathrm{U}_k$ is defined as 
\begin{eqnarray}\label{eqn:achievable rate}
    \mathcal{R}_{C,k} =  \kappa \log[2]{1+\gamma_{C,k}},
\end{eqnarray}
where $\kappa = (\tau_c - \tau_p)/\tau_c$ and $\gamma_{C,k}$ is the SINR of $\mathrm{U}_k$, and defined using \eqref{eqn:kth user recieved signal 2} as
\begin{eqnarray}\label{eqn:SINR}
    \gamma_{C,k} =  \frac{|\q{f}_{k}^{\mathrm{H}}\q{w}_k|^2}{\sum_{ j = 0, j\neq k}^{K} |\q{f}_{k}^{\mathrm{H}}\q{w}_j|^2 + 1}.
\end{eqnarray}

\subsection{Sensing Rate}\label{sec:SCNR}
This will be used as the metric for evaluating sensing performance. Although transmit beampattern gain and its mean squared error (MSE) are also used for this purpose, they overlook the effects of the receiver's beam pattern and clutter interference, which can cause detection ambiguities due to multi-path reflections \cite{mao2024, Lian2023, Hua2023}. Similarly, the CRLB focuses solely on the lower bound of estimation error, offering a measure of accuracy for parameters such as angle, distance, or velocity \cite{Ren2024, xia2025, Hua2024}. However, CRLB does not account for the sensing rate.  

Due to these limitations, the sensing rate is widely used  \cite{perovic2025, Behdad2024}. It quantifies the amount of sensing information captured (in bps/Hz), representing both the efficiency and quality of the sensing process. A higher sensing rate generally corresponds to a better sensing SCNR, leading to improved target detection and reduced estimation errors \cite{Behdad2024, choi2024, lu2025, rivetti2024}. While it does not directly measure localization error, the sensing rate is correlated with estimation accuracy and offers a communication-compatible, tractable metric for assessing sensing performance in integrated systems. Additionally, it supports a unified optimization of C\&S under shared resource constraints, making it particularly suitable for dynamic, information-driven applications, such as autonomous systems and surveillance, where rapid and high-quality sensing is essential \cite{Behdad2024, choi2024, lu2025, rivetti2024, Liu2022, luong2025}.

From \eqref{eqn:received echo 3}, the sensing rate at $\mathrm{RAP}_j$ is given by 
\begin{eqnarray}\label{eqn:sensing rate}
    \mathcal{R}_{S,j} =  \kappa \log[2]{1+\gamma_{S,j}},
\end{eqnarray}
where $\gamma_{S,j}$ is the SCNR at $\mathrm{RAP}_j$ and given as
\begin{eqnarray}\label{eqn:SCNR}
    \gamma_{S,j} &=&  \frac{\sum_{k=0}^{K} \Tr{\q{G}_{j,t} \q{w}_k \q{w}_k^{\mathrm{H}}  \q{G}_{j,t}^{\mathrm{H}}}}{\sum_{c=1}^{N_C} \sum_{k=0}^{K}\ \Tr{\q{G}_{j,c}  \q{w}_k \q{w}_k^{\mathrm{H}} \q{G}_{j,c}^{\mathrm{H}}} + M_R}.\quad
\end{eqnarray}

\begin{rem}\label{rem_rcs_target_clutter}
    {Although clutter may exhibit RCS values comparable to or even larger than that of the target, the proposed STCIB framework does not prioritize scatterers based on reflection magnitude (Section~{\ref{sec_DL_sol}}). Instead, beamforming is learned by maximizing the sensing-rate objective in {\eqref{eqn:sensing rate}}, which is governed by the SCNR. As shown in {\eqref{eqn:SCNR}}, target reflections contribute to the numerator, whereas clutter reflections, regardless of their RCS, appear in the denominator and thus degrade sensing performance. Consequently, the learning process penalizes beamforming patterns that enhance clutter interference and promotes spatial energy distributions that strengthen the target-reflected signal relative to clutter. The attention mechanism, therefore, captures feature interactions most relevant to optimizing the SCNR, rather than acting as an amplitude-based selector of strong reflectors {\cite{lee2019}}.}
\end{rem}

\section{Problem Formulation} \label{sec:problem formulation}Here, CF-ISAC beamforming is formulated under the SC, CC, and joint design. These help quantify the C\&S trade-offs under different priority settings. To this end, a unified problem formulation can be given as follows:
\begin{subequations}
\begin{eqnarray}\label{eqn_unified_prob}
       \!\!\!\!\!\! \mathcal P:~&& \max_{\q{W} } \quad \eta \sum_{k = 1}^{K} \log[2]{1+\gamma_{C,k}} \nonumber \\
        && \quad \quad \quad \quad + (1-\eta) \sum_{j = 1}^{N_R} \log[2]{1+\gamma_{S,j}}, \label{eqn_up_obj} \\
        \text{s.t}  && \beta_S \kappa \log[2]{1+\gamma_{S,j}} \geq \zeta_{\rm{th}}, ~\forall j , \label{eqn_up_sen_const} \\ 
        &&  \beta_C \kappa \log[2]{1+\gamma_{C,k}} \ge \vartheta_{\rm{th}}, ~\forall k , \label{eqn_up_comm_const} \\ 
        &&   \Vert\q{W}\Vert^2_F \leq \rho N_T, \label{eqn_up_power}
\end{eqnarray}
\end{subequations}
where $\eta$ is the weighting factor that defines the priorities for C\&S tasks, and $\{\beta_S, \beta_C\}$ are binary variables that activate the corresponding C\&S requirements. In particular,
\begin{enumerate}[label=\roman*)]
    \item $\{\eta = 0, \beta_C = 1, \beta_S = 0\}$ for SC design; this specific problem is denoted by  $\mathcal P_1$, 
    \item $\{\eta = 1, \beta_C = 0, \beta_S = 1\}$ for CC design; this specific problem is denoted by  $\mathcal P_2$, and
    \item $\{0<\eta<1, \beta_C = 1, \beta_S = 1\}$ for joint design;  this specific problem is denoted by  $\mathcal P_3$.
\end{enumerate}
Moreover, \eqref{eqn_up_sen_const} guarantees that each RAP meets the minimum required sensing rate, where $\zeta_{\rm{th}}$ denotes the predefined sensing rate threshold, and \eqref{eqn_up_comm_const} ensures that each user satisfies the minimum required communication rate, with $\vartheta_{\rm{th}}$ being the required communication rate threshold. Finally, \eqref{eqn_up_power} enforces the transmit power constraint at the TAPs, where $\rho$ denotes the transmit SNR of each TAP.

\section{Proposed Set Transformer-based Solution}\label{sec_DL_sol}
Here, we propose STCIB, an unsupervised Transformer-based learning method for $\mathcal{P}_1$, $\mathcal{P}_2$, and $\mathcal{P}_3$. It learns the input-output mapping by optimizing a custom loss function, eliminating the need for labeled training data \cite{qi2024}. While the classical Transformer architecture uses self-attention to capture dependencies across ordered sequences \cite{vaswani2017}, the ordering of elements is irrelevant in our setting. Therefore, we adopt the ST architecture of \cite{lee2019}, which processes unordered sets and ensures permutation invariance.
\begin{figure*}[!t]
    \centering
    \def\svgwidth{0.9\textwidth} 
    \fontsize{9}{4}\selectfont
    \graphicspath{{Figures/}}
    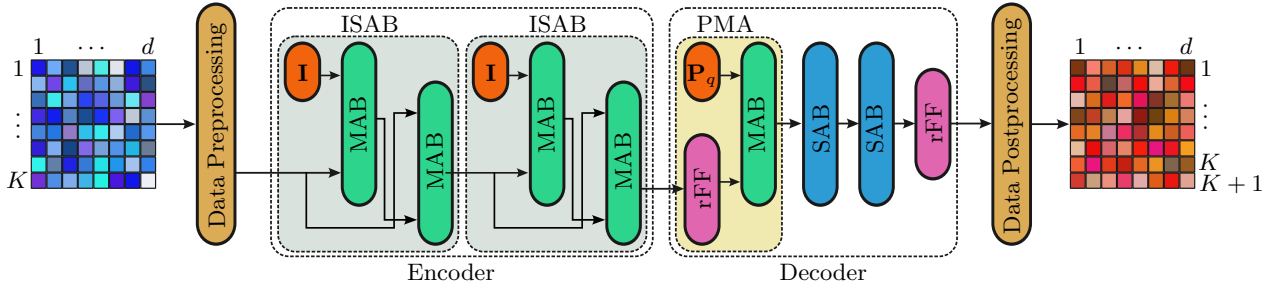
    \caption{The proposed ST-based architecture for unified beamformer designs in the CF-ISAC system.}
    \label{fig_proposed_DL_model}\vspace{-3mm}
\end{figure*}

Note that the beamforming solutions are obtained based on the estimated channels. Hence, from this point onward, the channel estimates obtained during the uplink channel estimation phase (see Section~\ref{subsec:Uplink_channel_estimation}) are employed in place of the true channels for the optimization problems. Accordingly, we denote $\hat{\q{f}}_k = [\hat{\q{h}}_{1k}^{\mathrm{H}},  \cdots, \hat{\q{h}}_{N_Tk}^{\mathrm{H}}]^{\mathrm{H}} \in \mathbb{C}^{N_TM_T \times 1}$.

\subsection{Data Pre-Processing}
The input to {STCIB} is the estimated communication user channels matrix $[\hat{\q{f}}_1, \cdots, \hat{\q{f}}_K]$. However, due to the inability of existing Transformer-based networks to process a complex input matrix, we use the I/Q transformation to separate the in-phase component $[\Re(\hat{\q{f}}_1), \cdots, \Re(\hat{\q{f}}_K)]$ and the quadrature component $[\Im(\hat{\q{f}}_1), \cdots, \Im(\hat{\q{f}}_K)]$ from the complex channel matrix, where $\Re(.)$ and $\Im(.)$ denote the real part and the imaginary part, respectively. Thus, the feature input matrix $\bar{\q{F}} \in \mathbb{R}^{K \times d}$ of the Transformer-based network can be defined by combining the in-phase and quadrature components as
\begin{equation}\label{eqn_model_input_matrix}
    \bar{\q{F}} = \bigg[\Big[\Re(\hat{\q{f}}_1^\mathrm{T}), \Im(\hat{\q{f}}_1^\mathrm{T})\Big]^{\mathrm{T}}, \cdots, \Big[\Re(\hat{\q{f}}_K^\mathrm{T}), \Im(\hat{\q{f}}_K^\mathrm{T})\Big]^{\mathrm{T}}\bigg]^{\mathrm{T}},
\end{equation}
where $d = 2 N_T M_T$. Next, layer normalization is applied to ensure stable training, i.e., 
\begin{equation}\label{eqn_input_norm}
    \tilde{\q{F}} = \mathrm{LayerNorm}(\bar{\q{F}}).
\end{equation}
Finally, the resultant matrix $\tilde{\q{F}}$ is projected onto higher $d$-dimensional vectors before passing the channel information into the model as
\begin{equation}\label{eqn_input_proj}
    \q{F} = \mathrm{rFF}(\tilde{\q{F}}),
\end{equation}
where the operator $\mathrm{rFF}(\cdot)$ is a row-wise feed forward (rFF) layer. Moreover, the primary computational cost of a single forward pass comes from the data pre-processing layer, with complexity $\mathcal{O}(KN_TM_T)$.

\subsection{ST-based Model Structure}
Based on the underlying principles of transformer architectures, our ST-based network effectively models the interactions between the user channels to generate beamformers without relying on the input order of the user channels due to its permutation-invariant attention mechanism \cite{lee2019}. It comprises an encoder and a decoder, both dependent on attention mechanisms. Our encoder consists of two induced set attention blocks (ISABs), which can be given as
\begin{equation}\label{eqn_encoder}
    {\rm{Encoder}}(\q{F}) = {\rm{ISAB}}_n({\rm{ISAB}}_n(\q{F})).
\end{equation}

\begin{rem}
    {The attention mechanism can be interpreted from a geometric learning perspective {\cite{lee2019}}. In CF-ISAC systems, the input feature set $\q{F}$ contains channel and system parameters of APs, users, and sensing targets. Although these features are high-dimensional, their feasible configurations are governed by physical propagation constraints such as geometry, pathloss, and angular relationships, causing them to lie on a structured low-dimensional manifold embedded in the ambient feature space {\cite{lee2019}}.}

    {The ST learns a permutation-invariant mapping that embeds this unordered set into a latent representation capturing the intrinsic manifold/low-dimensional structure {\cite{lee2019, Bengio2013}}. Through self-attention, each element aggregates information from all others, enabling the model to learn pairwise and higher-order interactions that characterize propagation relationships. This learned representation allows STCIB to approximate the mapping from system states to beamforming solutions without explicitly solving the underlying optimization problem {\cite{lee2019, Bengio2013}}.}
\end{rem}

The ISAB mechanism is described as follows:
\subsubsection{ISAB} As illustrated in Fig. \ref{fig_proposed_DL_model}, in addition to the input set $\q{F} \in \mathbb{R}^{K \times d}$, $n$ $d$-dimensional vectors $\q{I} \in \mathbb{R}^{n \times d}$ are defined within the ISAB itself, referred to as inducing points. They are trainable parameters that can also be trained with other parameters of the network. The ISAB consists of two multihead attention blocks (MABs), and the functionality of the ISAB with $n$ inducing points can be defined as
\begin{equation}\label{eqn_ISAB_function}
    {\rm{ISAB}}_n(\q{F}) = {\rm{MAB}}(\q{F}, {\rm{MAB}}(\q{I}, \q{F})).
\end{equation}
In \eqref{eqn_ISAB_function}, the inducing points $\q{I}$ are transformed via the first MAB by attending to the input set, i.e., aggregating all relevant information from the input elements, resulting in $n$ elements. Then, the transformed inducing points produce a set of $K$ elements by again attending to the input set. Here, in order to obtain meaningful features for the final objective, $\q{F}$ is first mapped to a lower-dimensional representation, which is subsequently reconstructed to generate the output.

\begin{figure}[!t]\vspace{-2mm}
\centering 
    \def\svgwidth{0.42\textwidth} 
    \fontsize{7}{4}\selectfont 
    \graphicspath{{Figures/}}
    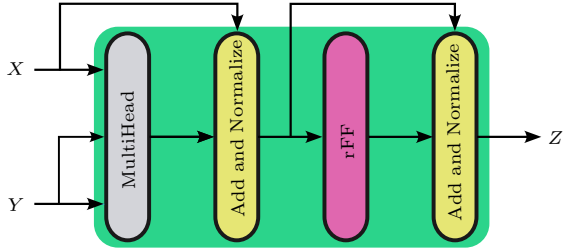\vspace{-2mm}
    \caption{MAB structure.}
    \label{fig_MAB}\vspace{-3mm}
\end{figure}

\subsubsection{MAB} The MAB is derived from the Transformer encoder architecture without positional encoding and dropout mechanisms to accommodate permutation-invariant set input. As shown in Fig. \ref{fig_MAB}, the functionality of the MAB can be defined as
\begin{eqnarray}\label{eqn_MAB}
    {\rm{MAB}}(\q{X}, \q{Y}) = {\rm{LayerNorm}}(\q{J} + {\rm{rFF}}(\q{J})), \label{eqn_MAP_function} 
\end{eqnarray}
where
\begin{eqnarray}
    \q{J} = {\rm{LayerNorm}}(\q{X} + {\rm{MultiHead}}(\q{X}, \q{Y}, \q{Y})). \label{eqn_J_function}
\end{eqnarray}
Here, $\q{X}, \q{Y} \in \mathbb{R}^{l_1 \times l_2}$ are two sets of $l_2$-dimensional vectors. In \eqref{eqn_J_function}, the function ${\rm{MultiHead}}(\cdot)$ is defined as
\begin{eqnarray}\label{eqn_MultiHead}
    {\rm{MultiHead}}(\q{Q}, \q{K}, \q{V}) = {\rm{Concat}}(\q{H}^{(1)}, \cdots, \q{H}^{(h)}) \q{W}^O, \label{eqn_MH_function}
\end{eqnarray}
where 
\begin{eqnarray}\label{eqn_head_function}
    \q{H}^{(i)} = {\rm{Attention}}\Big(\q{Q}^{(i)}, \q{K}^{(i)}, \q{V}^{(i)}\Big),\; \forall i, 
\end{eqnarray}
and $\q{Q}^{(i)} = \q{Q} \q{W}_i^Q \in \mathbb{R}^{m_1 \times \bar{d}_1}$, $\q{K}^{(i)} = \q{K} \q{W}_i^K \in \mathbb{R}^{m_2 \times \bar{d}_1}$, and $\q{V}^{(i)} = \q{V} \q{W}_i^K \in \mathbb{R}^{m_2 \times \bar{d}_2}$ are projections of the matrices $\q{Q} \in \mathbb{R}^{m_1 \times d_1}$ (query), $\q{K} \in \mathbb{R}^{m_2 \times d_2}$ (key), and $\q{V} \in \mathbb{R}^{m_2 \times d_3}$ (value) onto the $\bar{d}_1$, $\bar{d}_1$, and $\bar{d}_2$-dimensional vectors, respectively. Furthermore, $\q{W}_i^Q \in \mathbb{R}^{d_1 \times \bar{d}_1}$, $\q{W}_i^K \in \mathbb{R}^{d_2 \times \bar{d}_1}$ and $\q{W}_i^V\in \mathbb{R}^{d_3 \times \bar{d}_2}$ in \eqref{eqn_head_function}, and $\q{W}^O\in \mathbb{R}^{h\bar{d}_2 \times d_O}$ in \eqref{eqn_MH_function} are learnable parameters, $h$ is the number of heads, $d_1 = d_2$, $\bar{d}_1 = d_1/h$, $\bar{d}_2 = d_3/h$, and $d_O = d_1$. Unless stated otherwise, we use $d_1 = d_2 = d_3 = d$ and $\bar{d}_1 = \bar{d}_2 = d/h$. Moreover, the function ${\rm{Attention}}(\cdot)$ is defined as
\begin{equation}\label{eqn_Attention}
    {\rm{Attention}}\Big(\q{Q}^{(i)}, \q{K}^{(i)}, \q{V}^{(i)}\Big) = {\rm{softmax}}\Bigg(\frac{\q{Q}^{(i)} (\q{K}^{(i)})^{\mathrm{T}}}{\sqrt{\bar{d}_1}}\Bigg) \q{V}^{(i)},
\end{equation}
where the similarity between $\q{Q}^{(i)}$ and $\q{K}^{(i)}$ is captured by using their pairwise dot product with the weights generated by the softmax activation function.
    Furthermore, the computational cost is dominated by the ISAB block, with a complexity of $\mathcal{O}(Kn)$ per forward pass \cite{lee2019}.

Next, the final feature set $\q{U} \in \mathbb{R}^{K \times d}$ constructed by the encoder is passed through the decoder that aggregates encoded feature set by using pooling by multihead attention (PMA), applies the set attention block (SAB) followed by rFF to capture the interactions between the aggregated feature vectors. The decoding process can be described as
\begin{equation}\label{eqn_decoder}
    {\rm{Decoder}}(\q{U}) = {\rm{rFF}}({\rm{SAB}}({\rm{SAB}}({\rm{PMA}}_{K + 1}(\q{U})))),
\end{equation}
where the mechanisms of PMA and SAB are described below.

\subsubsection{PMA} In conventional set-based architectures, static pooling functions such as mean or average pooling are used in feature aggregation, providing permutation invariance \cite{zaheer2017}. Although these methods are simple and effective, they treat all inputs equally and are unable to capture the content-specific relevance. To circumvent this, PMA has been proposed to aggregate the feature-set $\q{U}$ generated by the encoder using a learnable set of $q$ seed vectors $\q{P}_q \in \mathbb{R}^{q \times d}$. The functionality of PMA is defined as
\begin{equation}\label{eqn_PMA}
    {\rm{PMA}}_q(\q{U}) = {\rm{MAB}}(\q{P}_q, {\rm{rFF}}(\q{U})).
\end{equation}
Note that since our model output requires $K + 1$ vectors for the users and the target, we set $q = K + 1$. In addition, the PMA block represents the primary computational bottleneck, with complexity of $\mathcal{O}(K^2)$ per forward pass.

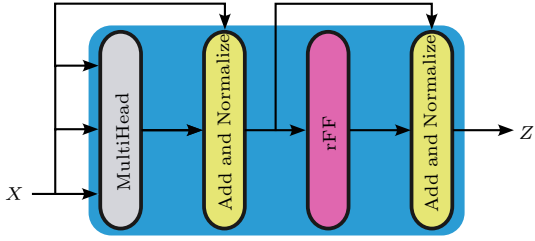
\begin{figure}[!t]\vspace{-2mm}
\centering 
    \def\svgwidth{0.42\textwidth} 
    \fontsize{7}{4}\selectfont 
    \graphicspath{{Figures/}}
\begingroup%
  \makeatletter%
  \providecommand\color[2][]{%
    \errmessage{(Inkscape) Color is used for the text in Inkscape, but the package 'color.sty' is not loaded}%
    \renewcommand\color[2][]{}%
  }%
  \providecommand\transparent[1]{%
    \errmessage{(Inkscape) Transparency is used (non-zero) for the text in Inkscape, but the package 'transparent.sty' is not loaded}%
    \renewcommand\transparent[1]{}%
  }%
  \providecommand\rotatebox[2]{#2}%
  \newcommand*\fsize{\dimexpr\f@size pt\relax}%
  \newcommand*\lineheight[1]{\fontsize{\fsize}{#1\fsize}\selectfont}%
  \ifx\svgwidth\undefined%
    \setlength{\unitlength}{383.88371505bp}%
    \ifx\svgscale\undefined%
      \relax%
    \else%
      \setlength{\unitlength}{\unitlength * \real{\svgscale}}%
    \fi%
  \else%
    \setlength{\unitlength}{\svgwidth}%
  \fi%
  \global\let\svgwidth\undefined%
  \global\let\svgscale\undefined%
  \makeatother%
  \begin{picture}(1,0.48599347)%
    \lineheight{1}%
    \setlength\tabcolsep{0pt}%
    \put(0,0){\includegraphics[width=\unitlength,page=1]{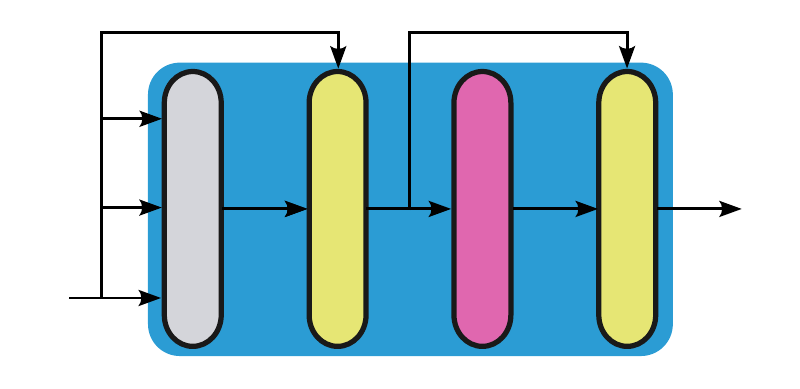}}%
    \put(0.25742869,0.11928923){\color[rgb]{0,0,0}\rotatebox{90.95068519}{\makebox(0,0)[lt]{\lineheight{1.25}\smash{\begin{tabular}[t]{l}$\mathrm{MultiHead}$\end{tabular}}}}}%
    \put(0.43513674,0.08319294){\color[rgb]{0,0,0}\rotatebox{90.95068519}{\makebox(0,0)[lt]{\lineheight{1.25}\smash{\begin{tabular}[t]{l}$\mathrm{Add\;and\;Normalize}$\end{tabular}}}}}%
    \put(0.7951437,0.08132988){\color[rgb]{0,0,0}\rotatebox{90.95068519}{\makebox(0,0)[lt]{\lineheight{1.25}\smash{\begin{tabular}[t]{l}$\mathrm{Add\;and\;Normalize}$\end{tabular}}}}}%
    \put(0.616694,0.18866019){\color[rgb]{0,0,0}\rotatebox{90.95068519}{\makebox(0,0)[lt]{\lineheight{1.25}\smash{\begin{tabular}[t]{l}$\mathrm{rFF}$\end{tabular}}}}}%
    \put(0.03921947,0.1027337){\color[rgb]{0,0,0}\rotatebox{0.95068519}{\makebox(0,0)[lt]{\lineheight{1.25}\smash{\begin{tabular}[t]{l}$X$\end{tabular}}}}}%
    \put(0.93574462,0.2095132){\color[rgb]{0,0,0}\rotatebox{0.95068519}{\makebox(0,0)[lt]{\lineheight{1.25}\smash{\begin{tabular}[t]{l}$Z$\end{tabular}}}}}%
  \end{picture}%
\endgroup%
\vspace{-2mm}
    \caption{SAB structure.}
    \label{fig_SAB}\vspace{-3mm}
\end{figure}

\subsubsection{SAB} The SAB is applied on the PMA output $\q{V} \in \mathbb{R}^{(K + 1) \times d}$ to further model the relationships among the $K + 1$ vectors. The SAB is illustrated in Fig.~\ref{fig_SAB}, and is defined using the MAB as
\begin{equation}\label{eqn_SAB}
    {\rm{SAB}}(\q{V}) = {\rm{MAB}}(\q{V}, \q{V}).
\end{equation}
The computational complexity of a forward pass is dominated by the SAB layer, which has a complexity of $\mathcal{O}(K^2)$ \cite{lee2019}. Moreover, the final rFF block in the decoder dominates the computational complexity of $\mathcal{O}(K N_T M_T)$ during each forward pass.

{The architectural configurations of STCIB are selected through preliminary design exploration guided by the Set Transformer model in {\cite{lee2019}}. Further investigation of alternative configurations can be treated in future work.}

\begin{rem}
    Both set operations SAB and ISAB are permutation equivariant, i.e., ${\rm{SAB}}(\q{V})$ and ${\rm{ISAB}}_n(\q{F})$ are permutation equivariant \cite{lee2019}. However, the computational complexity of the SAB scales quadratically with $K$, i.e., $\mathcal{O}(K^2)$, which is too expensive for a large-scale implementation. To circumvent this, ISAB has been proposed by introducing additional $m$ trainable vectors called inducing points that improve scalability while maintaining expressiveness. Then, the computational complexity of ${\rm{ISAB}}_n(\q{F})$ can be defined as $\mathcal{O}(Kn)$.
\end{rem}

{Note that the mapping learned by the ST is a low-dimensional manifold underlying the relationships among the APs, users, targets, and channel parameters governed by physical propagation laws. Heterogeneous large-scale fading conditions, reflected by different pathloss exponents, alter the relative channel strengths among network entities but do not modify the fundamental geometric relationships dictated by propagation physics. Such variations, therefore, change the distribution of data samples on the underlying manifold rather than its intrinsic structure or dimensionality {\cite{lee2019, Bengio2013}}. Consequently, the ST can still learn this manifold effectively provided that the training data captures representative heterogeneous channel conditions. Assumptions on large-scale fading parameters thus influence data statistics but do not simplify the learned manifold itself {\cite{Diluka2025}}.}

\subsection{Data Post-Processing} \label{sub_sec_post_processing}
The final output of our proposed model is the optimized beamforming matrix $\q{W}$. However, since the model cannot directly produce a complex-valued matrix, it returns the combination of the in-phase component $\Re(\q{W})$ and the quadrature component $\Im(\q{W})$, i.e., $\tilde{\q{W}} = [\Re(\q{W}^{\mathrm{T}}), \Im(\q{W}^{\mathrm{T}})] \in \mathbb{C}^{(K + 1) \times d}$. In order to generate $\tilde{\q{W}}$, the output of the Transformer-based network $\q{Z} \in \mathbb{R}^{K \times d}$ is passed through a custom Lambda layer to satisfy the transmit power constraint defined in \eqref{eqn_up_power}. In particular, the function of the Lambda layer is expressed as
\begin{eqnarray}\label{eqn_lambda_layer_fn}
    \tilde{\q{W}}=
    \begin{cases}
        \q{Z},& \Vert \q{Z} \Vert^2_F \leq \rho N_T,\\
        \sqrt{\rho N_T} \boldsymbol{Z}/ \Vert \boldsymbol{Z} \Vert_F, & \Vert \q{Z} \Vert^2_F > \rho N_T,
    \end{cases}
\end{eqnarray}
where $\Vert\cdot\Vert_F$ denotes the Frobenius norm. Based on the Lambda layer function, the total power of the TAPs is scaled to $[0, \rho N_T]$. Thus, the output of our proposed transformer-based model always meets the constraint, i.e., $\Vert \q{W} \Vert_F^2 < \rho N_T$, via the custom Lambda layer implementation.

\subsection{Custom Loss Function}
Unlike supervised learning, our unsupervised learning-based model does not require labels for training. Hence, we adopt a custom loss function for our model to improve system performance. The custom loss functions for $\mathcal{P}_1$, $\mathcal{P}_2$, and $\mathcal{P}_3$ are defined as follows:

\subsubsection{SC Optimization} \label{loss_sensing_centric} The power constraint for $\mathcal{P}_1$ is addressed by the Lambda layer in the post-processing stage. Hence, the custom loss function for $\mathcal{P}_1$ can be defined by using the penalty method as \cite{chang2022}
\begin{eqnarray}\label{eqn_sensing_custom_loss_fn}
    {\rm{Loss}}_{\mathcal{P}_1} &=& - \frac{1}{N_s} \sum_{n = 1}^{N_s} \Bigg[\sum_{j = 1}^{N_R} \mathcal{R}_{S,j}^{(n)} \nonumber \\
    && - \sum_{k = 1}^{K} \xi_k \Big(\min\{0, \mathcal{R}_{C,k}^{(n)} - \vartheta_{\rm{th}}\}\Big)^2 \Bigg],  
\end{eqnarray}
where $N_s$ is the number of samples, $\xi_k$ is the penalty parameter for the $k$-th user to determine the penalty term magnitude, and $\mathcal{R}_{C,k}^{(n)}$ and $\mathcal{R}_{S,j}^{(n)}$ are the communication rate of the $k$-th user and the sensing rate at $\mathrm{RAP}_j$, respectively, for the $n$-th sample.

\subsubsection{CC Optimization} Following a similar approach to Section \ref{loss_sensing_centric}, the custom loss function for $\mathcal{P}_2$ can be defined as
\begin{eqnarray}\label{eqn_communication_custom_loss_fn}
    {\rm{Loss}}_{\mathcal{P}_2} &=& - \frac{1}{N_s} \sum_{n = 1}^{N_s} \Bigg[\sum_{k = 1}^{K} \mathcal{R}_{C,k}^{(n)} \nonumber \\
    && - \sum_{j = 1}^{N_R} \varrho_j \Big(\min\{0, \mathcal{R}_{S,j}^{(n)} - \zeta_{\rm{th}}\}\Big)^2 \Bigg],  
\end{eqnarray}
where $\varrho_j$ is the penalty parameter for $\text{RAP}_j$.

\subsubsection{Joint  Optimization}
From $\mathcal{P}_3$, we consider the negative objective in \eqref{eqn_up_obj} as the loss function, and its constraint is addressed by the Lambda layer in the post-processing stage. Hence, the custom loss function for $\mathcal{P}_3$ can be defined as \cite{qi2024}
\begin{equation} \label{eqn_joint_custom_loss_fn}
    {\rm{Loss}}_{\mathcal{P}_3} = - \frac{1}{N_s} \! \sum_{n = 1}^{N_s} \!\Bigg[\eta \!\sum_{k = 1}^{K} \!\mathcal{R}_{C,k}^{(n)}  + (1-\eta)\! \sum_{j = 1}^{N_R} \!\mathcal{R}_{S,j}^{(n)} \Bigg].
\end{equation}

\subsection{Offline Training and Online Deployment}
Our proposed model is implemented using the PyTorch library, which automatically handles backpropagation \cite{paszke2017}. The dataset required for training and deployment is generated by stacking the channel realizations obtained from \eqref{eqn:Channel_h_ik} via Monte Carlo simulations. The feature input matrix $\q{F}$ is constructed according to \eqref{eqn_model_input_matrix}. {We train the model using mini-batches and update its parameters with the Adam optimizer to minimize the custom loss function {\cite{diederik2015}}.} Early stopping is employed to prevent overfitting. 

{Training is performed offline under a fixed system configuration. Once training is complete, the learned parameters are fixed, and the model is deployed online to generate the beamforming matrix for new channel realizations via a single forward pass. Because STCIB is permutation-invariant, reordering APs or users does not affect the output. However, since the number of PMA seed vectors determines the output dimensionality, the model is trained and deployed for a fixed number of users and streams. If the network size changes, retraining or fine-tuning becomes necessary. In practice, however, CF network configurations evolve much more slowly than channel conditions {\cite{elrashidy2024, demirhan2024, lee2019}}. Moreover, conventional optimization-based methods such as CCPA and ALM-MO also require updates when system configurations change {\cite{zargari2025, zargari2024}}, making retraining or adaptation a common requirement in such scenarios.}

{\textbf{Algorithm~{\ref{alg_stcib}}} provides the STCIB framework for optimizing AP beamforming through offline training and online deployment, where $N_B$ is the number of batches.}

\begin{algorithm}[!t] 
\caption{: {STCIB framework}}
\label{alg_stcib}
\begin{algorithmic}[1]
\STATE \textbf{Offline Training Phase}
\STATE \textbf{Input:} $[\hat{\mathbf{f}}_1,\ldots,\hat{\mathbf{f}}_K]$, 
$\mathbf{G}_{j,m}$ for $m\in\{t,c\}$, $\rho$, rate threshold/weighting factors  ($\zeta_{\rm th}$ for SC, $\vartheta_{\rm th}$ for CC, $\eta$ for joint design), penalty parameters ($\xi_k\,\forall k$ for SC, $\varrho_j\,\forall j$ for CC).
\STATE \textbf{Output:} Optimal beamforming matrix $\mathbf{W}^\ast$.
\STATE \textbf{Initialize} network parameters.
\REPEAT
\FOR{$b = \{1,\ldots,N_B\}$}
\STATE \textbf{Preprocessing:}
Perform I/Q transformation to obtain $\bar{\mathbf{F}}$, then, compute $\mathbf{F} = \text{rFF}(\text{LayerNorm}(\bar{\mathbf{F}}))$.
\STATE \textbf{Encoder-decoder mapping:} Compute $\mathbf{Z} = \text{Decoder}(\text{Encoder}(\mathbf{F}))$.
\STATE \textbf{Postprocessing:}
Obtain $\tilde{\mathbf{W}}$ using \eqref{eqn_lambda_layer_fn}, followed by
I/Q transformation to produce $\mathbf{W}^\ast$
\STATE \textbf{Loss computation:}
Evaluate the custom loss function (i.e., \eqref{eqn_sensing_custom_loss_fn} for SC, \eqref{eqn_communication_custom_loss_fn} for CC, and \eqref{eqn_joint_custom_loss_fn} for joint design).
\STATE \textbf{Parameter update:}
Compute gradients and update network parameters using the Adam optimizer.
\ENDFOR
\UNTIL{convergence}.
\STATE \textbf{Online Deployment Phase}
\STATE \textbf{Input:} $[\hat{\mathbf{f}}_1,\ldots,\hat{\mathbf{f}}_K]$.
\STATE \textbf{Output:} Beamforming matrix $\mathbf{W}^\ast$.
\STATE Perform forward propagation (preprocessign, encoder-decoder mapping, postprocessing) using the trained model to obtain $\mathbf{W}^\ast$.
\end{algorithmic} 
\end{algorithm}

\subsection{Computational Complexity} \label{subsec_STCIB_complexity}
{By analyzing each individual block, the computational complexities for a single pass in data pre-processing, encoder, and decoder layers are $\mathcal{O}(K N_T M_T)$, $\mathcal{O}(K n)$, and $\mathcal{O}(K N_T M_T + K^2)$, respectively {\cite{lee2019}}. The computational cost of the proposed STCIB framework naturally divides into two stages:}
\begin{enumerate}
    \item {\textit{Offline training phase}: The model parameters are learned using a large dataset through iterative optimization of the loss function {\cite{chang2022}}. While computationally intensive, it is performed only once prior to deployment and can be accelerated using parallel hardware. This training complexity is $\mathcal{O}(I_{\rm{max}} L (K N_T M_T + K^2 + K n))$, where $I_{\rm{max}}$ and $L$ are the maximum number of iterations and the number of training samples, respectively. These reflect the costs of feature processing, attention operations, and feedforward layers.}

    \item {\textit{Online deployment}: During this phase, beamforming vectors are generated through a single forward pass without iterative optimization. The corresponding inference complexity is $\mathcal{O}(K N_T M_T + K^2 + K n)$.}
\end{enumerate}
{Once trained offline, the model is reused for multiple online instances under a fixed system configuration (e.g., fixed $N_T, N_R, M_T, M_R, K$). For each new channel realization within this configuration, beamforming solutions are generated directly through inference without retraining.}

\section{Localization}
We use the 2D MUltiple SIgnal Classification (MUSIC) algorithm to evaluate the sensing performance at azimuth and elevation angles \cite{Schmidt1986}. We consider $\tau = \tau_c - \tau_p$ snapshots available at each RAP. The MUSIC algorithm is defined based on the sample covariance matrix of the received signal at RAP \cite{Schmidt1986}, i.e., 
\begin{equation} \label{eqn_MUSIC_cov}
    R_j = \frac{1}{\tau} \sum_{t = 1}^{\tau} \q{y}_j^{(t)} \q{y}_j^{(t) \mathrm{H}},
\end{equation}
where $\q{y}_j^{(t)}$ is the $t$-th snapshot of the received signal at $\mathrm{RAP}_j$. The noise subspace matrix $\q{N}_j \in \mathbb{C}^{M_R \times (M_R - (N_C + 1))}$ for the given target and clutters is constructed from the eigenvectors of $R_j$ corresponding to its $M_R - (N_C + 1)$ smallest eigenvalues. The 2D MUSIC spectrum at $\mathrm{RAP}_j$ is then defined as \cite{Schmidt1986}
\begin{equation} \label{eqn_MUSIC_spec}
    \Omega_j(x, y) = \frac{1}{\q{a}_R^{\mathrm{H}}(x, y) \q{N}_j \q{N}_j^{\mathrm{H}} \q{a}_R(x, y)},
\end{equation}
where $x$ and $y$ are azimuth and elevation angles at a point in the 2D grid. In this spectrum, the peaks indicate the presence of the target or clutters, which can be located via a grid search.

To further quantify localization accuracy, we define the localization root MSE (RMSE) based on true angles of the target and the estimated angles from the grid search through the MUSIC spectrum. The localization RMSE is given as 
\begin{equation} \label{eqn_localization_RMSE}
    \mathrm{RMSE} = \sqrt{(\psi^{\mathrm{true}} - \psi^{\mathrm{est}})^2 + (\theta^{\mathrm{true}} - \theta^{\mathrm{est}})^2},
\end{equation}
where $\{\psi^{\mathrm{true}},  \theta^{\mathrm{true}}\}$ and $\{\psi^{\mathrm{est}}, \theta^{\mathrm{est}}\}$ are the actual and estimated azimuth/elevation angles of the target, respectively.

\section{Numerical Results}
Next, we present the numerical results to evaluate the efficiency of the proposed Transformer-based framework.

\subsection{Benchmarking}\label{sec_benchmark}
The following benchmark schemes are employed to evaluate the performance of the proposed STCIB approach.
\subsubsection{CCPA Beamforming} This method is based on the standard SDR and SCA techniques. In particular, we first redefine the beamforming variables as $\q{W}_k \triangleq \q{w}_k \q{w}_k^{\mathrm{H}}$ for $k \in \{0, 1, \cdots, K\}$, where $\q{W}_k \in \mathbb{C}^{N_T M_T \times N_T M_T}$ is a semi-definite matrix, i.e., $\q{W}_k  \succeq 0$, with $\text{Rank}(\q{W}_k) = 1$. Thereby, $\mathcal{P}_1$,  $\mathcal{P}_2$, and $\mathcal{P}_3$ are reformulated into standard SDPs by relaxing the rank one constraints \cite{sidiropoulos2006}. The Gaussian randomization technique is used to recover the rank one solution \cite{Luo2010}.

\subsubsection{ALM-MO Beamforming} This approach addresses problems $\mathcal{P}_1$, $\mathcal{P}_2$, and $\mathcal{P}_3$ within the ALM-MO framework proposed in  \cite{zargari2025} (Appendix A). In particular, the non-convex sum-log objective is transformed using the FP technique \cite[\textit{Theorem 3}]{shen2018a}, while all constraints other than the transmit power limits at the TAPs are incorporated through the ALM formulation \cite{liu2020b, zargari2024}. Furthermore, the inherent geometric structure of the TAP beamforming vectors, characterized by the Frobenius-norm constraint $\Vert\q{W}\Vert^2_F \leq \rho N_T$, is exploited to reformulate the problem as an optimization over a Riemannian manifold. The resulting manifold-constrained problem is then efficiently solved using the ALM-MO algorithm introduced in \cite{zargari2025, zargari2024}.

{The implementation details and parameter settings of the CCPA and ALM-MO benchmarks follow the configurations reported in {\cite{zargari2025, zargari2024}}. In particular, the number of Gaussian randomization trials, solver tolerances, and stopping criteria are adopted directly from these references.}

\subsubsection{CNN-based Beamforming} This re-models the STCIB architecture in Fig. \ref{fig_proposed_DL_model}, by changing the encoder and decoder structures. The encoder is re-defined by using two convolutional layers to extract local spatial features without altering the size of the grid, providing output $\q{U} \in \mathbb{R}^{K \times d}$ for the input $\q{F} \in \mathbb{R}^{K \times d}$.  The decoder is replaced by a transposed convolution, which is a common learnable upsampling layer in CNNs, followed by another convolutional layer \cite{dumoulin2016, elrashidy2024}. This transposed convolutional layer adds one extra row, which is generated based on nearby features, resulting in network output $\q{Z} \in \mathbb{R}^{K+1 \times d}$ for the encoder output $\q{U}$ \cite{elrashidy2024}.

{For the conventional optimization benchmarks (CCPA and ALM-MO), the beamforming variables are initialized using a standard random feasible strategy for each channel realization. Specifically, the initial beamforming matrix is randomly generated and scaled to satisfy the transmit power constraint {\cite{zargari2025}}. In each algorithm, the iterative solvers terminate when either the relative improvement in the objective value falls below a predefined tolerance or when a maximum number of iterations is reached. These stopping criteria and tolerance settings are kept identical across all experiments and are consistent with those used in the performance evaluations.}

\subsection{Simulation Setup and Parameters} \label{subsec_simulation_setup}
Unless otherwise stated, the simulation and model parameters are given in Table \ref{tab_sim_params}. Furthermore, pathloss is modeled as {$PL(r) = L_0 + 10 \alpha \mathrm{log}(r)$}, where $\alpha = 3$ is the pathloss exponent, $L_0 = -40$ dB is the pathloss at the reference distance $1$ m, and {$r$} is the distance in meters. The Rician factors $K_{ik}\;\forall i, k$ are randomly drawn from the set $[1, 3]$. {The values of $\tau_c$ and $\tau_p$ are {\num{196}} and $K = 4$, respectively, resulting in $\tau = \tau_c - \tau_p = 192$. Moreover, the MUSIC spectrum-based localization grid spans from {\qty{-100}{\degree}} to {\qty{100}{\degree}} in both azimuth and elevation, with a resolution of {\qty{0.5}{\degree}} in each direction.}

\begin{table}[t]
    \centering
    \caption{Simulation and model parameters}
    \label{tab_sim_params}
    \begin{tabular}{c c c c}
        \hline
        \textbf{Parameter}   &  \textbf{Value} & \textbf{Parameter} & \textbf{Value} \\ \hline \hline
        $N_T$   &   \num{8} & Training set size & \num{50000} \\
        $N_R$   &   \num{4} & Validation set size & \num{25000} \\
        $M_T$   &   \num{4} & Prediction set size & \num{25000} \\
        $M_R$   &   \num{4} & Learning rate & \num{1e-4} \\
        $K$     &   \num{4} & Maximum epoch size & \num{500} \\
        $N_C$   &   \num{2} & Mini-batch size & \num{1000} \\
        $\rho$  &   \qty{10}{\dB} & Patience & \num{20} \\ \hline
    \end{tabular}
    \vspace{-3mm}
\end{table}

{For training, validation, and prediction of the proposed framework, $1 \times 10^5$ samples are generated using Monte Carlo simulations. During the offline training stage, we allocate $50\%$ and $25\%$ of the total samples for training and validation, respectively. The remaining $25\%$ is used for the beamformer prediction during the online deployment.}

{All experiments are conducted on a  Ubuntu 24.04 workstation. It has an Intel\textsuperscript{\textregistered} Core\textsuperscript{TM} i9-14900K CPU (32 cores) and 64 GB of RAM, as well as a single NVIDIA GeForce RTX 5070 Ti GPU with 16 GB of VRAM. All STCIB models are implemented in Python 3.10 using PyTorch 2.9.0 and trained on the GPU with CUDA 12.8. However, the online deployment of these models is performed without GPU support.} In addition,  CCPA and ALM-MO are implemented in MATLAB 2024b on the same machine without GPU acceleration. Therefore, this identical hardware configuration ensures a fair comparison of computational complexity across all algorithms (see Section \ref{subsec_computational_complexity}). 

\begin{figure}[!t]
    \centering
    \includegraphics[width=0.8\linewidth]{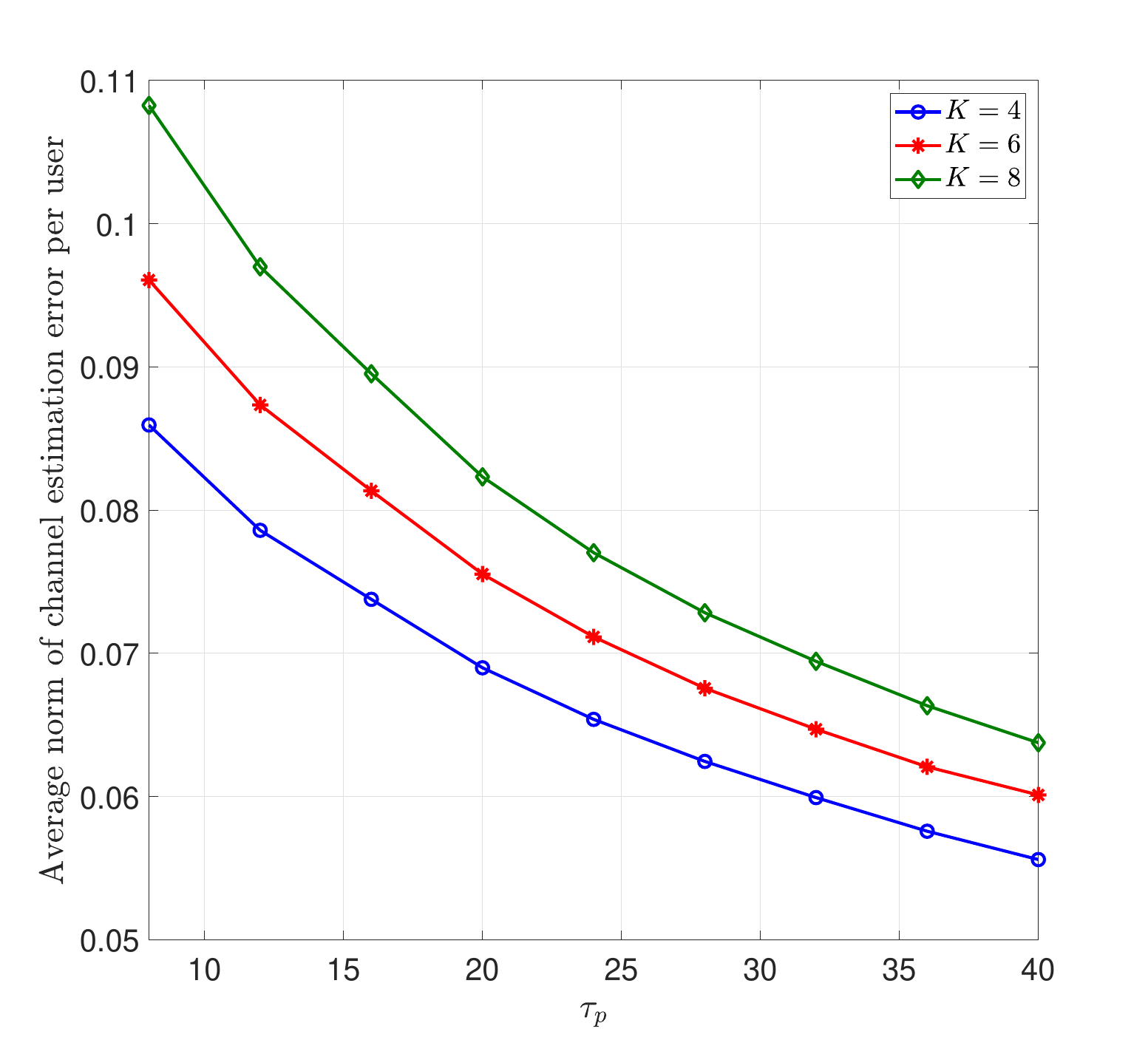}\vspace{-1mm}
    \caption{Channel estimation error.} 
    \label{fig_estimation_error}\vspace{-3mm}
\end{figure}

\subsection{Channel Estimation}
Fig.~\ref{fig_estimation_error} depicts the relationship between the average estimation error norm per user and the number of pilot samples $(\tau_p)$ for $K = \{4, 6, 8\}$. Increasing  $\tau_p$ reduces the estimation error, since a longer pilot interval provides more accurate channel information. For example, when $K = \num{4}$, the average estimation error norm per user decreases by \qty{35.35}{\percent} as $\tau_p$ increases from \num{8} to \num{40}. On the other hand, for a fixed $\tau_p$, the pilot length is shared among all users. As $K$ increases, each user receives fewer pilots within the $\tau_p$ interval, leading to less accurate channel estimation and higher estimation error. For instance, at $\tau_p = \num{20}$, increasing $K$ from \num{4} to \num{6} and \num{8} results in a \qty{9.48}{\percent} and \qty{19.33}{\percent} increase, respectively, in the average estimation error norm per user.

\begin{figure}[!t]
    \centering
    \includegraphics[width=0.8\linewidth]{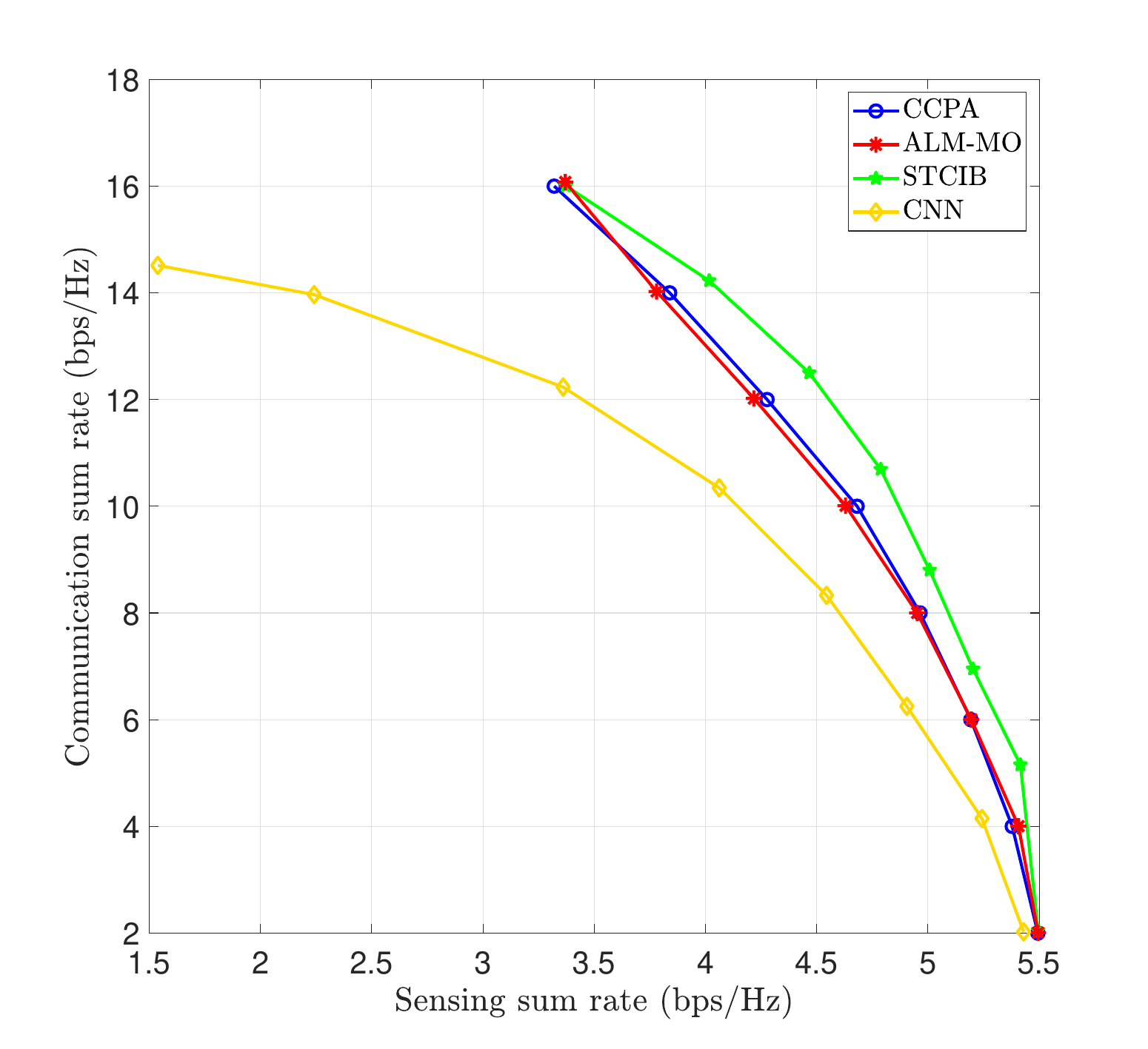}\vspace{-1mm}
    \caption{SC design trade-off.}
    \label{fig_sensing_tradeoff}\vspace{-3mm}
\end{figure}

\begin{figure}[!t]
    \centering
    \includegraphics[width=0.8\linewidth]{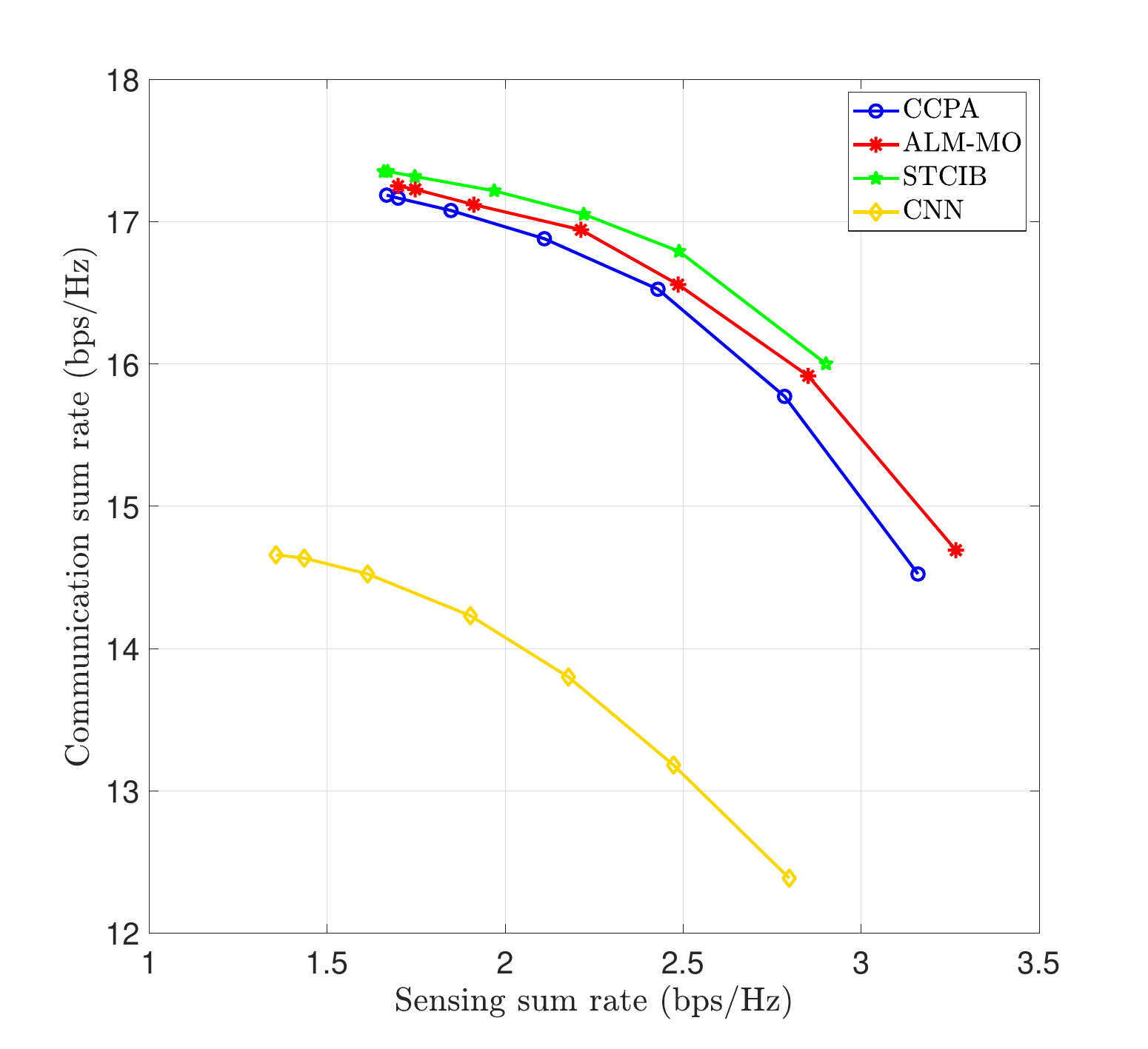}\vspace{-1mm}
    \caption{CC design trade-off.}
    \label{fig_communication_tradeoff}\vspace{-3mm}
\end{figure}

\subsection{Communication-Sensing Trade-offs}
Since both C\&S functions share the same resources in ISAC, there is a trade-off between these performances. The tradeoff curve may be described by the pair $(R_s, R_c)$ where $R_c$ and $R_s$ are the C\&S sum rates.  This section demonstrates that {STCIB} achieves a better $(R_c, R_s)$ than other algorithms.  Fig.~\ref{fig_sensing_tradeoff} and Fig.~\ref{fig_communication_tradeoff} investigate this for SC and CC designs, respectively.

Fig.~\ref{fig_sensing_tradeoff} is generated by varying $\vartheta_{\rm{th}}$ from \qty{0.5}{bps/\Hz} to \qty{4.0}{bps/\Hz} in increments of \qty{0.5}{bps/\Hz} for $\mathcal{P}_1$. As $\vartheta_{\rm{th}}$ increases, the communication rate requirements consume more resources, thereby degrading the sensing performance. For instance, in STCIB, the sensing sum rate decreases by \qty{38.62}{\percent}, while the communication sum rate increases by \qty{680.54}{\percent} as $\vartheta_{\rm{th}}$ grows from \qty{0.5}{bps/\Hz} to \qty{4.0}{bps/\Hz}. Moreover, Fig.~\ref{fig_sensing_tradeoff} shows that STCIB consistently outperforms CCPA, {ALM-MO}, and CNN in solving $\mathcal{P}_1$. For example, at $\vartheta_{\rm{th}} = \qty{2.0}{bps/\Hz}$, STCIB improves the sensing sum rate by $\qty{0.87}{\percent}$, $\qty{1.13}{\percent}$, $\qty{10.20}{\percent}$ compared to CCPA, {ALM-MO}, and CNN, respectively. At the same threshold, STCIB also meets the communication rate requirement, achieving a communication sum rate of \qty{8.8}{bps/\Hz}, which exceeds the required sum rate of $K \times \vartheta_{\rm{th}} = 4 \times  \qty{2.0}{bps/\Hz} = \qty{8.0}{bps/\Hz}$.

Fig.~\ref{fig_communication_tradeoff} illustrates the trade-off performance for $\mathcal{P}_2$ as $\zeta_{\rm{th}}$ increases from \qty{0.1}{bps/\Hz} to \qty{0.7}{bps/\Hz} in steps of \qty{0.1}{bps/\Hz}. As $\zeta_{\rm{th}}$ increases, sensing consumes more resources, reducing communication performance. In STCIB, raising $\zeta_{\rm{th}}$ from \qty{0.1}{bps/\Hz} to \qty{0.7}{bps/\Hz} leads to a \qty{7.78}{\percent} decrease in the communication performance, while the sensing sum rate improves by \qty{74.90}{\percent}. As per Fig.~\ref{fig_communication_tradeoff}, STCIB consistently achieves a better tradeoff than  CCPA, {ALM-MO}, and CNN in solving $\mathcal{P}_2$. For example, at $\zeta_{\rm{th}} = \qty{0.4}{bps/\Hz}$, STCIB improves the communication sum rate by a \qty{2.00}{\percent}, \qty{1.62}{\percent}, and $\qty{20.99}{\percent}$ improvement relative to CCPA, {ALM-MO}, and CNN, respectively. At the same threshold, STCIB also satisfies the sensing rate requirement, achieving a sensing sum rate of \qty{1.969}{bps/\Hz}, which exceeds the required $N_R \times \zeta_{\rm{th}} = 4 \times \qty{0.4}{bps/\Hz} = \qty{1.6}{bps/\Hz}$.

STCIB outperforms CCPA and {ALM-MO} because these methods struggle with the non-convexity of $\mathcal{P}.$ CCPA depends on convex relaxations (e.g., SDR, SCA), which introduce approximation errors and require Gaussian randomization to recover rank-1 solutions, often yielding suboptimal or even infeasible results. {ALM-MO} methods are more flexible but rely on auxiliary variables and augmented constraints; their convergence is sensitive to parameter choices and initial conditions, making them prone to local optima. STCIB avoids these limitations by leveraging permutation-invariant self-attention in STs to learn nonlinear dependencies among APs, users, and targets, thereby capturing global interactions and balancing objectives more effectively. This yields consistently higher rates and a wider trade-off region. While the CNN baseline also avoids convex approximations, its fixed grid structure limits it to local patterns, thereby limiting its ability to model complex relationships, which explains STCIB's superior performance.

Due to space constraints, we focus on SC and CC designs. The joint ISAC design in $\mathcal{P}_3$ shows similar trends and comparable performance, so it is omitted for brevity.

\begin{figure}[!t]
    \centering
    \includegraphics[width=0.8\linewidth]{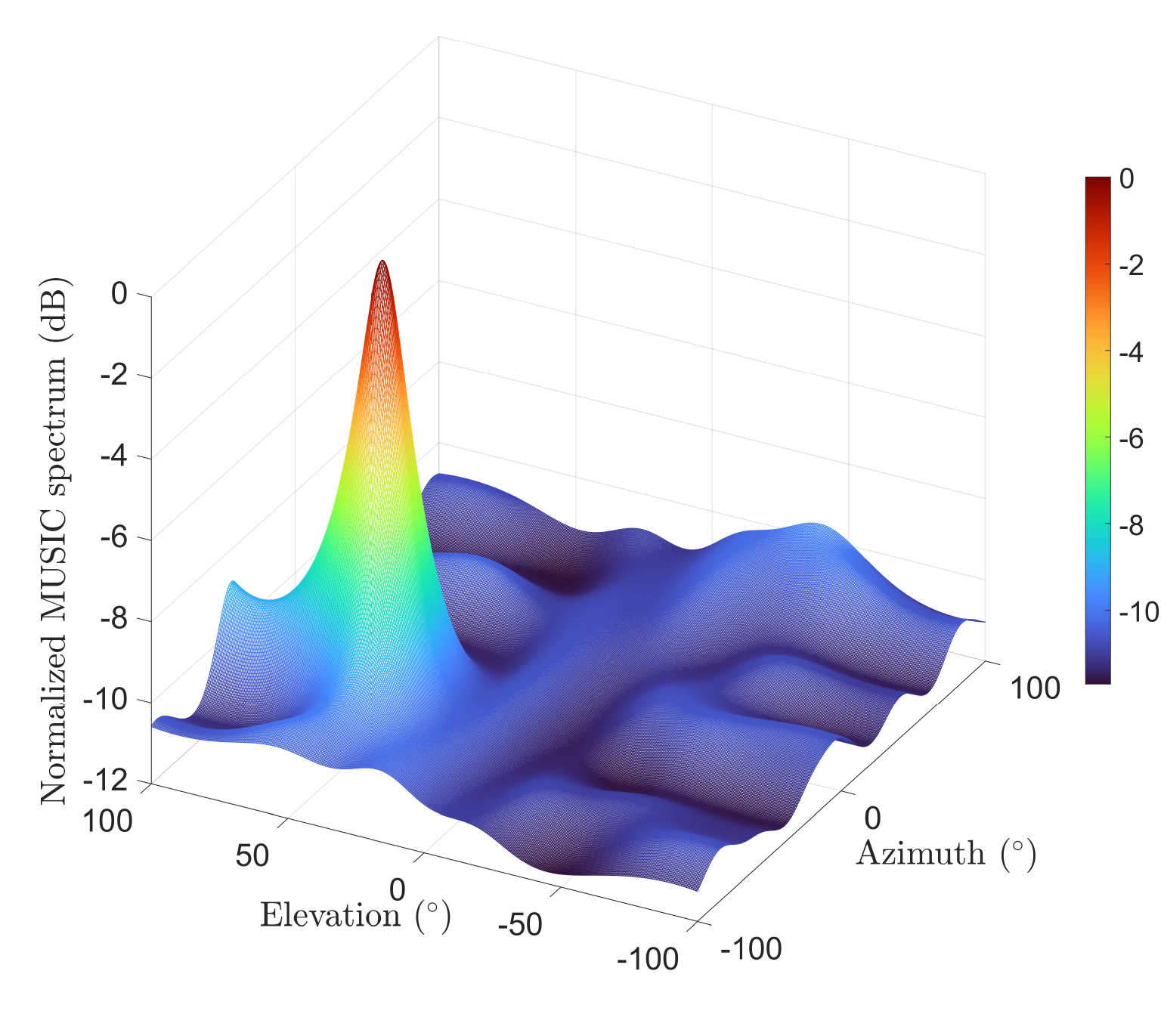}\vspace{-1mm}
    \caption{Joint ISAC design MUSIC spectrum $\eta = \num{0.2}$.}
    \label{fig_joint_music_2}\vspace{-3mm}
\end{figure}

\begin{figure}[!t]
    \centering
    \includegraphics[width=0.8\linewidth]{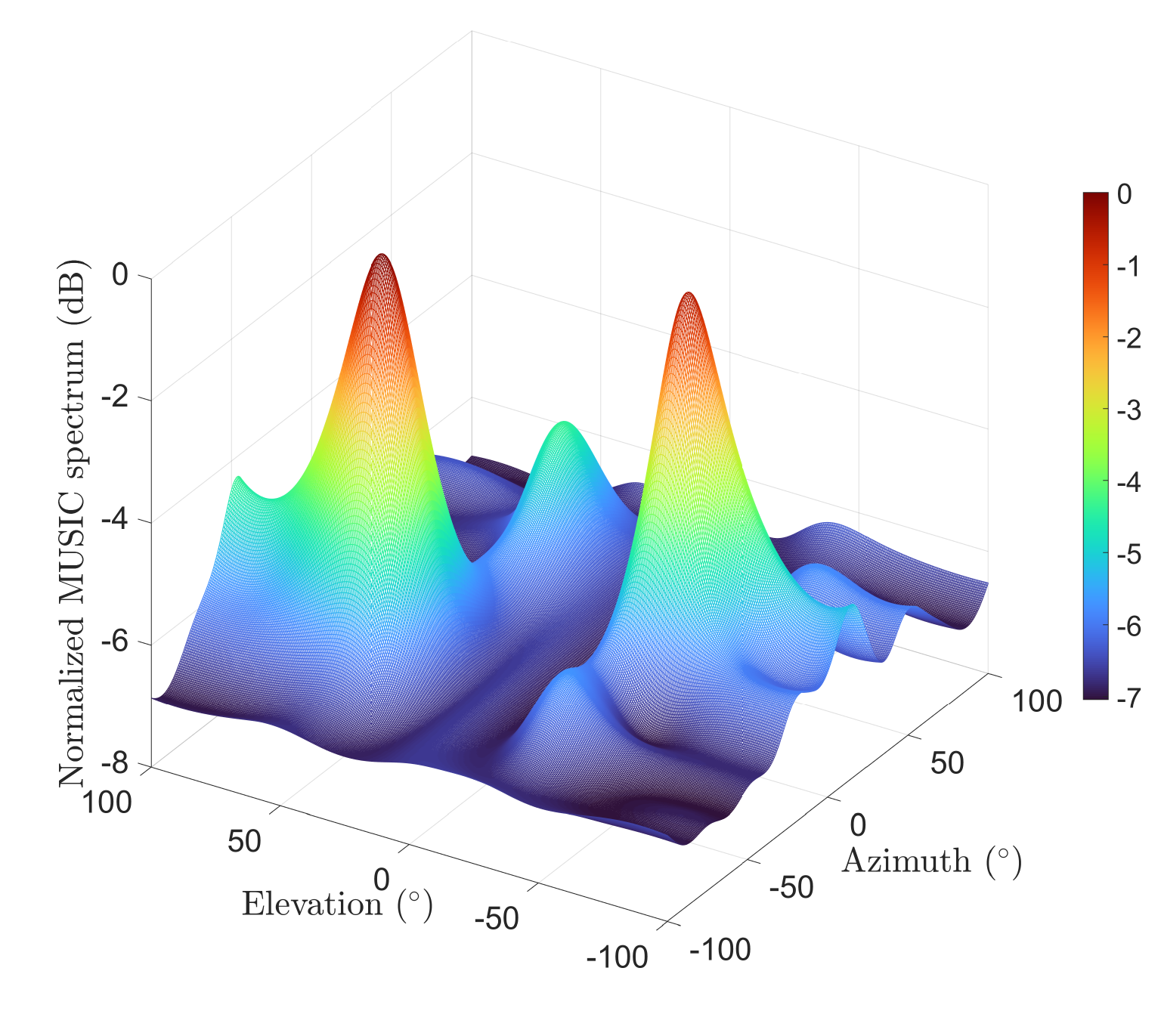}\vspace{-1mm}
    \caption{Joint ISAC design MUSIC spectrum $\eta = \num{0.8}$.}
    \label{fig_joint_music_8}\vspace{-3mm}
\end{figure}

\subsection{Localization Performance}
MUSIC  enhances ISAC-based localization by providing high-resolution angle-of-arrival estimation from the multi-antenna signals already used for C\&S. MUSIC can resolve multiple closely spaced targets and achieve much finer angular accuracy than conventional beamforming.  Fig.~\ref{fig_joint_music_2} and Fig.~\ref{fig_joint_music_8} depict the 2D MUSIC spectrum of STCIB at $\mathrm{RAP}_1$ for the joint design with $\eta = \num{0.2}$ and $\eta = \num{0.8}$, respectively. The target is placed at \qty{-45}{\degree} azimuth and \qty{45}{\degree} elevation relative to $\mathrm{RAP}_1$, while the clutters are  at \{\qty{45}{\degree}, \qty{0}{\degree}\} azimuth and \{\qty{30}{\degree}, \qty{-45}{\degree}\} elevation angles.

The spectral peaks indicate the locations of the target and clutter. As shown in Fig.~\ref{fig_joint_music_2} and Fig.~\ref{fig_joint_music_8}, increasing $\eta$ improves communication performance, whereas sensing performance declines, thereby reducing localization capability. Specifically, at the lower $\eta = \num{0.2}$ (Fig.~\ref{fig_joint_music_2}), the MUSIC spectrum exhibits a single peak corresponding to the target, effectively suppressing clutter interference. In contrast, at the higher $\eta = \num{0.8}$ (Fig.~\ref{fig_joint_music_8}), clutter interference increases, producing additional peaks that make target identification more difficult and degrade localization accuracy.

\begin{table}[t!]
    \centering
    \caption{Localization RMSE under different techniques.}
    \label{tab_rmse}
    \begin{tabular}{|p{2.7cm}|c|c|c|c|}
        \hline
        \textbf{Design \& Trade-off} & \textbf{CCPA} & \textbf{{ALM-MO}} & \textbf{CNN} & \textbf{STCIB} \\ \hline \hline
        
        SC ($\vartheta_{\rm{th}}$ = \qty{2}{bps/\Hz}) 
        & \qty{1.140}{^{\circ}} & \qty{0.967}{^{\circ}} & \qty{1.225}{^{\circ}} & \qty{0.615}{^{\circ}} \\ 
        
        SC ($\vartheta_{\rm{th}}$ = \qty{4}{bps/\Hz}) 
        & \qty{1.926}{^{\circ}} & \qty{1.895}{^{\circ}} & \qty{5.557}{^{\circ}} & \qty{1.392}{^{\circ}} \\ \hline
        
        CC ($\zeta_{\rm{th}}$ = \qty{0.2}{bps/\Hz}) 
        & \qty{1.082}{^{\circ}} & \qty{1.370}{^{\circ}} & \qty{3.088}{^{\circ}} & \qty{2.061}{^{\circ}} \\ 
        
        CC ($\zeta_{\rm{th}}$ = \qty{0.7}{bps/\Hz}) 
        & \qty{0.506}{^{\circ}} & \qty{0.774}{^{\circ}} & \qty{1.291}{^{\circ}} & \qty{1.000}{^{\circ}} \\ \hline
        
        Joint ($\eta$ = 0.2) 
        & \qty{0.981}{^{\circ}} & \qty{1.478}{^{\circ}} & \qty{2.316}{^{\circ}} & \qty{1.913}{^{\circ}} \\ 
        
        Joint ($\eta$ = 0.8) 
        & \qty{1.941}{^{\circ}} & \qty{3.118}{^{\circ}} & \qty{4.753}{^{\circ}} & \qty{3.636}{^{\circ}} \\ \hline
        
    \end{tabular}\vspace{-3mm}
\end{table}

To further evaluate localization performance, we give the localization RMSE \eqref{eqn_localization_RMSE} for different designs and solvers in Table \ref{tab_rmse}. In the SC design, STCIB outperforms CCPA, {ALM-MO}, and CNN, achieving a lower RMSE value. For example, in the SC design with $\vartheta_{\mathrm{th}} = \qty{2}{bps/\Hz}$, STCIB reduces RMSE by \qty{46.05}{\percent}, \qty{36.40}{\percent}, and \qty{49.80}{\percent}, compared to CCPA, {ALM-MO}, and CNN, respectively. In CC designs, increasing $\zeta_{\mathrm{th}}$ allocates more resources to sensing, improving localization; STCIB's RMSE decreases by \qty{51.48}{\percent} as $\zeta_{\mathrm{th}}$ increases from \qty{0.2}{bps/\Hz} to \qty{0.7}{bps/\Hz}.

\begin{figure}[!t]
    \centering
    \includegraphics[width=0.8\linewidth]{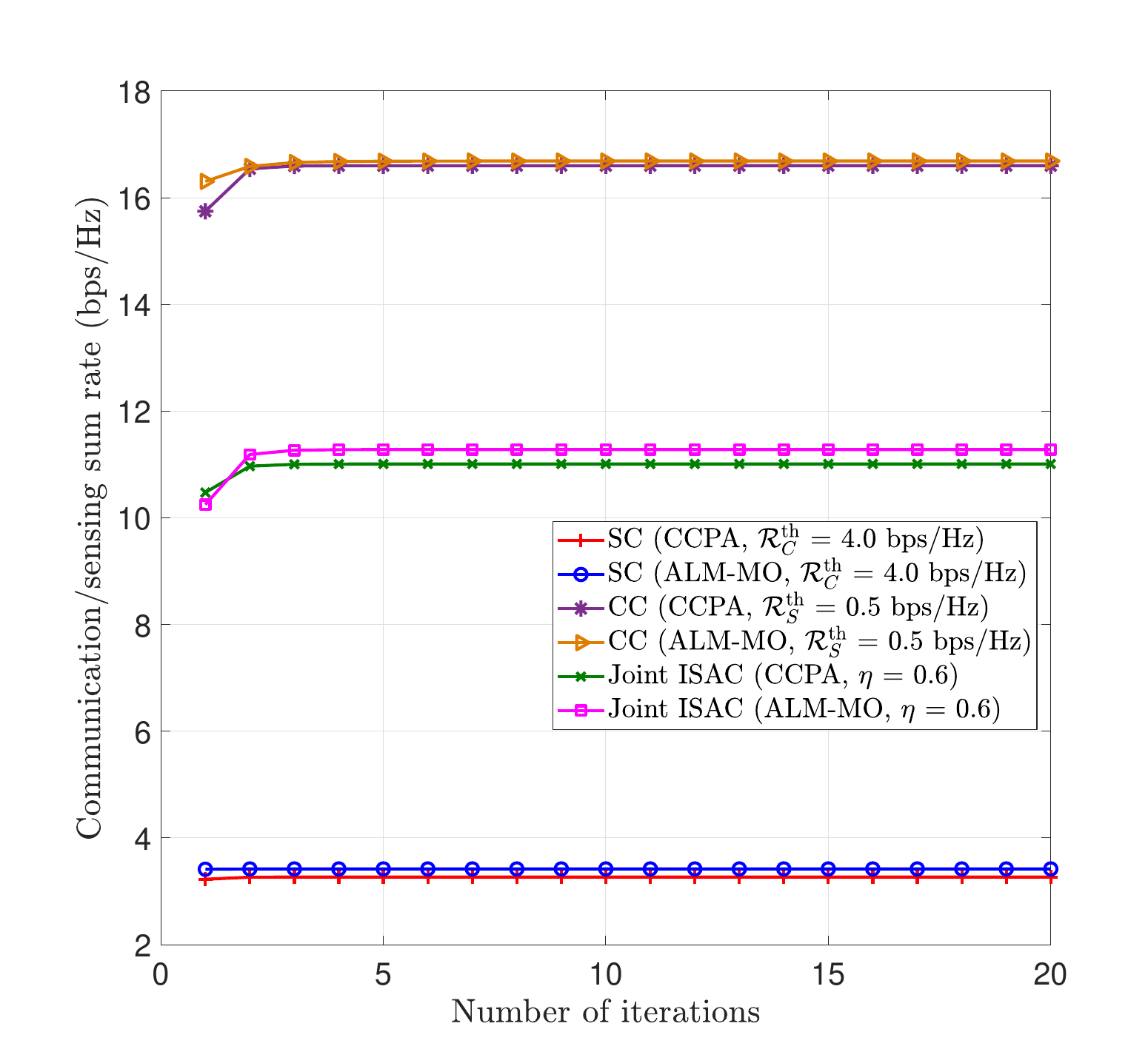}\vspace{-1mm}
    \caption{CCPA and {ALM-MO} objective function convergence.}
    \label{fig_ccpa_mo_convergence}\vspace{-3mm}
\end{figure}

\begin{figure}[!t]
    \centering
    \includegraphics[width=0.8\linewidth]{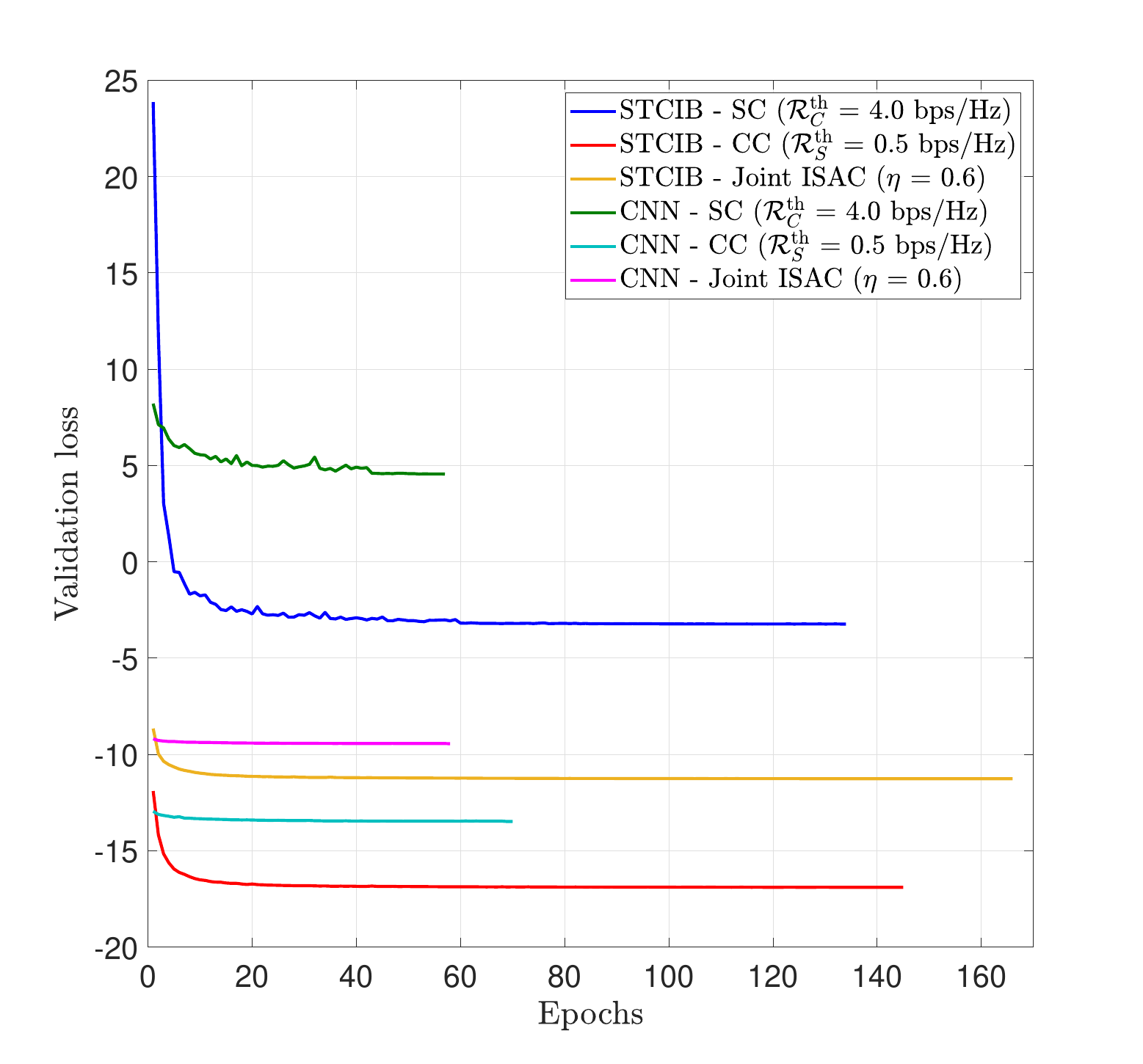}\vspace{-1mm}
    \caption{STCIB loss function convergence.}
    \label{fig_dl_loss_convergence}\vspace{-3mm}
\end{figure}

\subsection{Convergence Behavior}
Fig.~\ref{fig_ccpa_mo_convergence} illustrates the convergence of the CCPA and {ALM-MO} algorithms for $\mathcal{P}_1$-$\mathcal{P}_3$ as a function of the number of iterations, whereas Fig.~\ref{fig_dl_loss_convergence} shows the validation loss convergence of STCIB training over epochs. The loss functions for $\mathcal{P}_1$, $\mathcal{P}_2$, and $\mathcal{P}_3$ are defined in \eqref{eqn_sensing_custom_loss_fn}, \eqref{eqn_communication_custom_loss_fn}, and \eqref{eqn_joint_custom_loss_fn}, respectively. Although CCPA/{ALM-MO} are iterative optimization algorithms and STCIB is a learning-based model, convergence can be assessed by how efficiently each method converges to a stable objective value. From Fig.~\ref{fig_ccpa_mo_convergence}, both CCPA and {ALM-MO} converge within a few iterations, with the latter consistently achieving higher final objectives (e.g., \qty{4.70}{\percent}, \qty{0.53}{\percent}, and \qty{2.45}{\percent} improvements over CCPA in the SC, CC, and joint ISAC designs, respectively).  Similarly, Fig.~\ref{fig_dl_loss_convergence} shows that (STCIB, CNN) exhibits smooth and stable loss reduction during training and converges to ${\rm{Loss}}_{\mathcal{P}_1}=(\num{-3.230}, \num{4.563})$,  ${\rm{Loss}}_{\mathcal{P}_2}=(\num{-16.897}, \num{-13.484})$, and  ${\rm{Loss}}_{\mathcal{P}_3}=(\num{-11.257}, \num{-9.433})$ for the three designs in (\ref{eqn_sensing_custom_loss_fn}), and (\ref{eqn_communication_custom_loss_fn}), and (\ref{eqn_joint_custom_loss_fn}),  respectively.  Unlike CCPA and {ALM-MO}, which require iterative convergence for every new instance, STCIB and CNN learn the input-output mapping offline and provide near-instantaneous inference during deployment. Although the CNN converges faster than STCIB, it reaches higher loss values because it cannot capture global relationships or learn complex patterns, leading to lower performance.

Overall, although all the algorithms achieve reliable convergence,  STCIB offers a key advantage by shifting the computational burden offline and eliminating the need for online iterative optimization.

\begin{figure}[!t]
    \centering
    \includegraphics[width=0.8\linewidth]{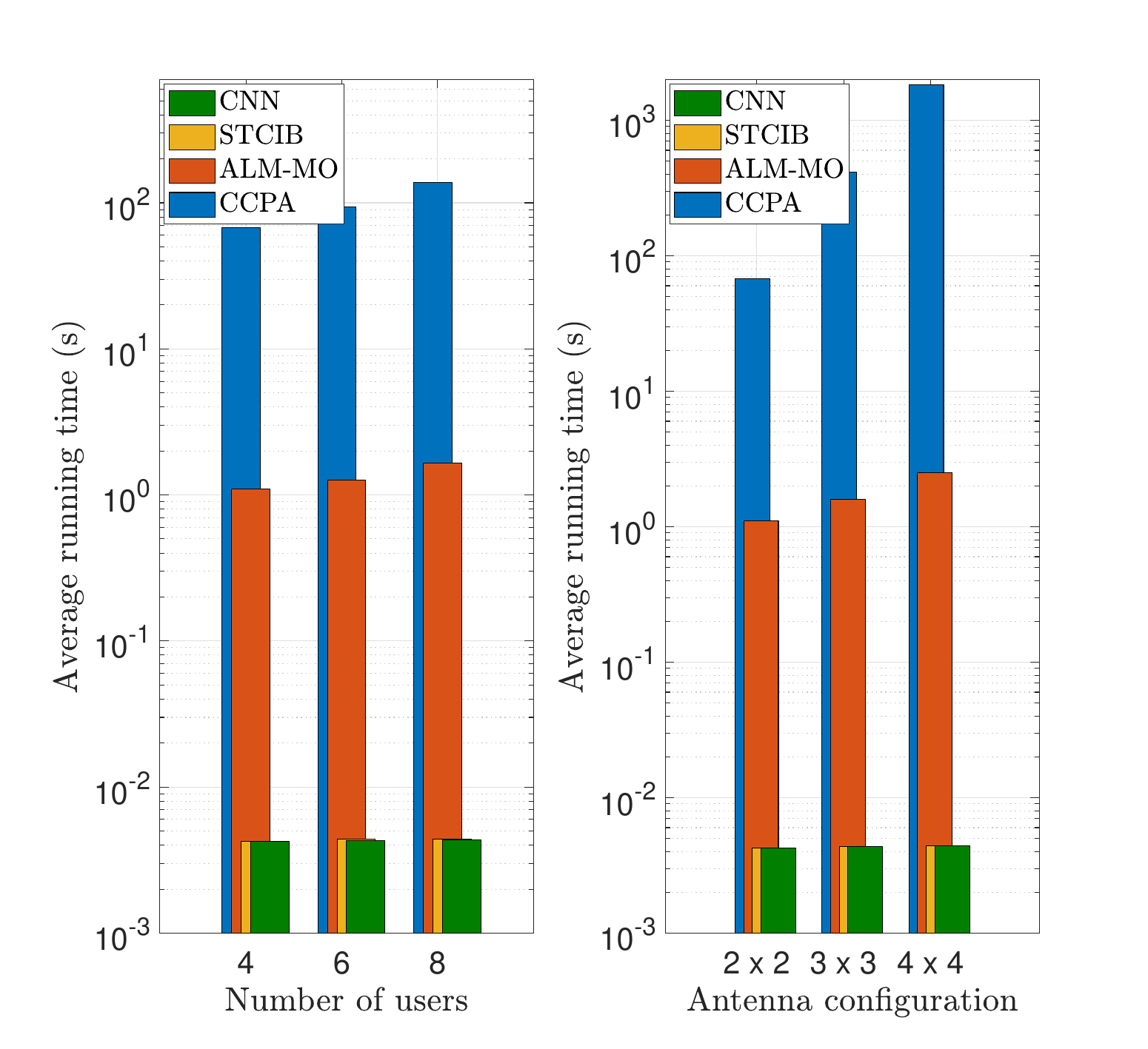}\vspace{-1mm}
    \caption{{Average online runtime of joint ISAC design with $\eta = 0.4$.}}
    \label{fig_joint_running_time}\vspace{-3mm}
\end{figure}

\subsection{Computation Complexity, Runtime, and Scalability} \label{subsec_computational_complexity}
{Let $\mathbb{E}(T)$ denote the average online runtime of any one of the four algorithms}. Note that for both STCIB and CNN,   $\mathbb{E}(T)$ is measured during their online operation. As described in \ref{subsec_simulation_setup}, all four algorithms are executed on the same hardware to ensure a fair comparison of computational complexity, runtime, and scalability (i.e., for an increasing number of users and antenna configurations).  

{For completeness, the computational burden of the benchmark methods can also be characterized analytically. For the CNN-based approach, the computational cost consists of both offline training and online deployment. The offline training complexity scales as $\mathcal{O}(I_{\rm{max}} L (K N_T M_T))$, while the online inference complexity is $\mathcal{O}(K N_T M_T)$, since beamforming solutions are generated through a single forward pass. In contrast, conventional optimization approaches solve iterative problems for each channel realization, leading to substantially higher per-instance complexities. Specifically, ALM-MO incurs a complexity of $\mathcal{O}(T(N_T M_T K + N_T M_T (K+N_R)^2))$, while CCPA requires $\mathcal{O}\left( \sqrt{N_TM_T} \log{\frac{1}{\epsilon}} (m(N_TM_T)^3 + m^2(N_TM_T)^2 + m^3) \right)$, where $T$ denotes the number of iterations, $\epsilon$ is solver accuracy, and $m$ is the number of constraints, i.e., $m=2K+1$ for SC design, $m=N_R+K+1$ for CC design, and $m=K+1$ for joint design {\cite{zargari2025}}.}

To this end, Fig.~\ref{fig_joint_running_time} presents $\mathbb{E}(T)$ for the joint ISAC design with $\eta = \num{0.4}$. Here, the left sub-plot shows $\mathbb{E}(T)$ as a function of the number of users, while the right sub-plot examines the impact of increasing the antenna configurations. 

As shown in Fig.~\ref{fig_joint_running_time}, STCIB dramatically reduces $\mathbb{E}(T)$ compared with CCPA and ALM-MO. For example, with four users, it only requires \qty{0.01}{\percent} and \qty{0.39}{\percent} runtime of that required by CCPA and ALM-MO, respectively. With six users, the runtime further decreases to merely \qty{0.0032}{\percent} and \qty{0.27}{\percent}, respectively. A similar trend is observed across antenna configurations. For a $2 \times 2$ configuration, the runtime of STCIB amounts to only \qty{0.01}{\percent} and \qty{0.39}{\percent} of that of CCPA and ALM-MO, respectively, while for the $4 \times 4$ configuration, it further decreases to just \qty{0.0003}{\percent} and \qty{0.18}{\percent}, respectively.

The $\mathbb{E}(T)$ data show that STCIB has dramatically lower complexity and superior scalability compared with CCPA and {ALM-MO}, i.e., as the number of users or antennas increases. {A similar trend is evident in both SC and CC designs (thus, omitted for brevity).} Conversely, STCIB moderately increases $\mathbb{E}(T)$ compared to CNN, but yields noticeably greater performance gains. For example, with four users, STCIB achieves \qty{14.82}{\percent} and \qty{31.60}{\percent} improvements in sensing and communication rates, respectively, while increasing $\mathbb{E}(T)$ by only \qty{0.26}{\percent} relative to CNN. 

{Note that the runtime comparison is intended to illustrate the difference in computational complexity between iterative optimization-based beamforming and feedforward inference of the proposed learning-based method. The comparison is not meant to evaluate MATLAB and PyTorch as software platforms.}

\subsection{{Feasibility Analysis}}
{In STCIB, the transmit power constraint in {\eqref{eqn_up_power}} is strictly enforced through the data post-processing stage (Section~{\ref{sub_sec_post_processing}}), whereas the minimum C\&S rate constraints are incorporated as penalty terms in the custom loss functions (see {\eqref{eqn_sensing_custom_loss_fn}} and {\eqref{eqn_communication_custom_loss_fn}}). As penalty-based methods may not guarantee strict feasibility for every channel realization, we perform an empirical feasibility evaluation on the test dataset, summarized in Table~{\ref{tab_feasibility}}.}

{The feasibility rate denotes the fraction of test samples satisfying the required rate constraints. The average violation measures the mean deviation from the threshold, while the worst-case violation captures the maximum deviation. Together, these metrics provide a comprehensive assessment of constraint satisfaction.}

{Results show that STCIB achieves feasibility rates exceeding {\qty{98}{\percent}} for both SC and CC designs. For example, in the SC design with $\vartheta_{\rm{th}}={\qty{3}{bps/\Hz}}$, the average and worst-case violations are {\qty{0.0940}{bps/\Hz}} and {\qty{0.3531}{bps/\Hz}}, corresponding to only {\qty{3.13}{\percent}} and {\qty{11.77}{\percent}} deviations, respectively. Similarly, in the CC design with $\zeta_{\rm{th}}={\qty{0.2}{bps/\Hz}}$, violations remain below {\qty{2.90}{\percent}} on average and {\qty{8.20}{\percent}} in the worst case. These small deviations confirm that STCIB reliably satisfies C\&S constraints while preserving computational efficiency.}

\begin{table}[t!]
    \centering
    \caption{{Feasibility results of the proposed STCIB framework.}}
    \label{tab_feasibility}
    \begin{tabular}{|p{2.7cm}|c|c|c|}
        \hline
        \textbf{Design \& Trade-off} & \textbf{\makecell{Feasibility \\ rate}} & \textbf{\makecell{Average \\ violation \\ (\qty{}{bps/\Hz})}} & \textbf{\makecell{Worst-case \\ violation \\ (\qty{}{bps/\Hz})}} \\ \hline \hline
        
        SC ($\vartheta_{\rm{th}}$ = \qty{1}{bps/\Hz}) 
        & \qty{99.00}{\percent} & \num{0.0043} & \num{0.2333}\\ 
        
        SC ($\vartheta_{\rm{th}}$ = \qty{3}{bps/\Hz}) 
        & \qty{98.50}{\percent} & \num{0.0940} & \num{0.3531}\\ \hline
        
        CC ($\zeta_{\rm{th}}$ = \qty{0.2}{bps/\Hz}) 
        &\qty{99.48}{\percent} & \num{0.0058} & \num{0.0164} \\ 
        
        CC ($\zeta_{\rm{th}}$ = \qty{0.4}{bps/\Hz}) 
        & \qty{99.06}{\percent} & \num{0.0244} & \num{0.0746}\\ \hline
        
    \end{tabular}
    \vspace{-3mm}
\end{table}

\subsection{{Robustness Against Channel Variation and CSI Impairments}}
\begin{figure}[!t]
    \centering  \vspace{-4mm}
    \includegraphics[width=0.8\linewidth]{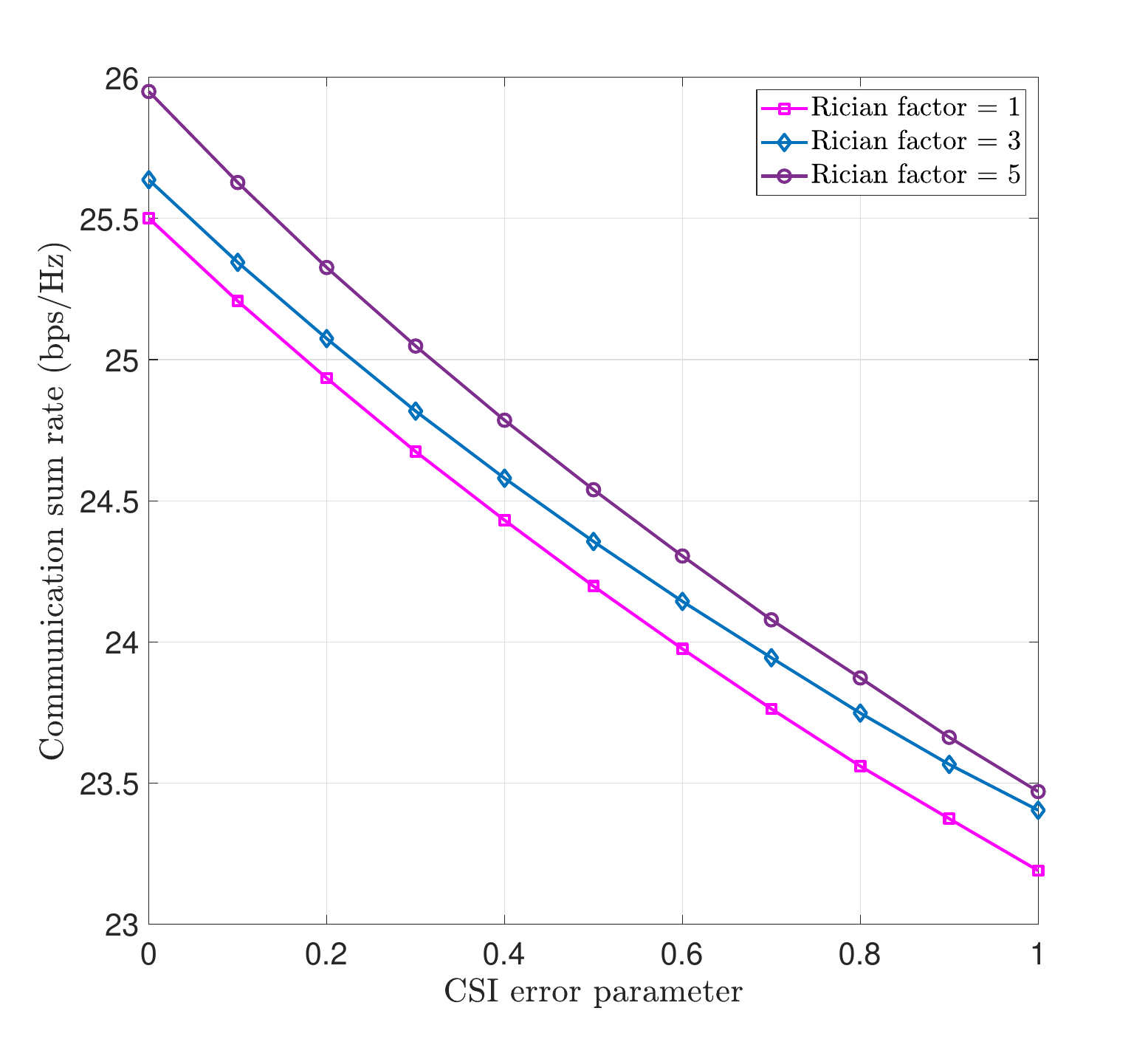}
    \caption{{Communication sum rate as a function of CSI errors for various Rician factors}. \vspace{-6mm}} 
    \label{fig_CSI_error} 
\end{figure}

{Fig.~{\ref{fig_CSI_error}} examines the impact of imperfect CSI and the Rician factor on communication rates. The CSI errors reduce beamforming accuracy, while higher Rician factors make the channels more deterministic and dominated by LoS components. The CSI errors are modeled as $\tilde{g} = g + \epsilon$, where $g \in \{[\q{h}_{ik}]_{m}\}$ for $m \in \{1, \cdots, M_T\}$ is the true channel, $\tilde{g}$ is the estimated channel, and $\epsilon \sim \mathcal{CN}(0, \sigma_{\epsilon}^2)$ is estimation noise with $\sigma_{\epsilon}^2 = \chi |g|^2$. Here, $|g|$ denotes the magnitude of the true channel and $0 \le \chi \le 1$ is the CSI error parameter that characterizes the estimation accuracy {\cite{Shengli2004,  Kay1993}}. The error variance $\sigma_{\epsilon}^2$ quantifies the channel estimation quality. The STCIB model is trained using perfect CSI $g$ ($\chi = 0$) for three different Rician factors of {\num{1}}, {\num{3}}, and {\num{5}}. This trained model is then used to infer beamformers for channels with estimation errors. The sum rates are then evaluated for the mismatched beamformers to examine the STCIB model's robustness, specifically for channels with dominant LoS components.}

{Fig.~{\ref{fig_CSI_error}} shows the communication rate versus $\chi$ for the CC design with $\zeta_{\rm{th}} = \qty{0.2}{bps/\Hz}$. The rate decreases as $\chi$ increases from {\num{0}} to {\num{1}} due to larger mismatches between actual and estimated channels. For example, at a Rician factor of {\num{1}}, the rate drops by {\qty{9.97}{\percent}}. Conversely, higher Rician factors improve the rate because of stronger LoS dominance. At $\chi = {\num{0.5}}$, the rate increases by {\qty{0.65}{\percent}} and {\qty{1.41}{\percent}} when the Rician factor rises from {\num{1}} to {\num{3}} and {\num{5}}, respectively. Overall, the results confirm the robustness of the proposed STCIB to CSI errors and channel variations as the rate loss due to channel mismatches is not significant.}

 \vspace{5mm}
\section{Conclusion} 
Two main computing paradigms exist for CF-ISAC beamforming: (i) classical iterative optimizers (e.g., CCPA and ALM-MO), which rely on approximations and relaxations and incur high computational cost; and (ii) learning-based approaches built on fixed-topology CNNs. Despite recent advances, both remain limited in their ability to model global dependencies and to scale efficiently. To address these gaps, we introduced STCIB, the first ST-based CF-ISAC beamforming framework. By leveraging permutation-invariant self-attention, STCIB captures long-range interactions among APs, users, and targets, and operates fully unsupervised using tailored loss functions, eliminating the need for labeled data. Across SC, CC, and joint ISAC designs, STCIB consistently surpasses the CNN baseline while maintaining comparable computational and time complexity. Moreover, it achieves orders-of-magnitude reductions ($\sim$\num{e3}) in computation and runtime relative to CCPA and ALM-MO, while slightly outperforming them in accuracy. Overall, the results demonstrate the robustness and strong effectiveness of the STCIB framework.

Building on the flexibility of STs, several promising research directions naturally follow. First, the global-attention mechanism can be extended to multi-target sensing and cooperative tracking, enabling STCIB to handle larger sets of spatially distributed scatterers, moving objects, and extended targets. Second, the permutation-invariant structure makes STCIB well-suited to scalable CF-massive MIMO systems, where the numbers of APs and users vary dynamically; incorporating adaptive tokens or hierarchical attention could enable real-time beamforming in ultra-dense networks. Moreover, because STs can process heterogeneous inputs, STCIB can evolve into a multi-task learning framework that jointly optimizes beamforming, localization, target-state prediction, and environment mapping. Finally, the unsupervised formulation positions STCIB as a strong candidate for online and continual learning, allowing adaptation to changing propagation conditions, mobility patterns, and sensing requirements without retraining from scratch. Advancing these directions may help establish attention-driven CF-ISAC as a foundational tool for next-generation intelligent wireless systems.


\section*{{Appendix A: ALM-MO Beamforming}} \label{APPENDIX_A}
{This section briefly outlines the ALM-MO technique employed in the benchmarking (i.e., Section~{\ref{sec_benchmark}}) scheme to effectively solve the optimization problems $\mathcal{P}_1 - \mathcal{P}_3$. For brevity, we only present the solution for $\mathcal{P}_1$. Nevertheless, a similar approach can be applied to solve $\mathcal{P}_2$ and $\mathcal{P}_3$.}

{Before proceeding to the solutions, we first introduce an index matrix as a $(K+1)$-order identity matrix $\q{E}$ to select each column of $\q{W}$, i.e., the $k$-th column of $\q{W}$ is given as $\q{w}_k = \q{W} \q{E}_k$, where $\q{E}_k$ is the $(k+1)$-th column of $\q{E}$. Thereby, the beamforming vectors in the SINR at $\mathrm{U}_k$ and the SCNR at $\mathrm{RAP}_j$ are replaced with  $\q{W}$ and $\q{E}$ {\cite{zargari2025}}.}

{To overcome the challenges in the sum-log objective in $\mathcal{P}_1$, we employ the Lagrangian dual transform to move $\gamma_{S,j}$ outside the log function. Then, the problem $\mathcal{P}_1$ can be reformulated as maximizing the objective function} $\bar{f}(\q{W}, \boldsymbol{\mu}_S) = \sum_{j = 1}^{N_R} \log[2]{1 + \mu_{S,j}} + \frac{1}{\text{ln}(2)}  \sum_{j = 1}^{N_R} \Big( - \mu_{S,j} + \frac{(1 + \mu_{S,j}) \gamma_{S, j}}{1 + \gamma_{S, j}} \Big)$, {where $\boldsymbol{\mu}_S = [\mu_{S,1}, \cdots, \mu_{S,N_R}]$ is the vector of auxiliary variables introduced by FP {\cite[\textit{Theorem 3}]{shen2018a}}. The resultant problem can be considered as a two-part optimization problem {\cite{shen2018a}}, i.e., (i) an outer optimization over $\q{W}$ for fixed $\boldsymbol{\mu}_S$ and (ii) an inner optimization over $\boldsymbol{\mu}_S$ for fixed $\q{W}$.}

{\textit{Optimizing $\boldsymbol{\mu}_S$ for fixed $\q{W}$:} For a fixed $\q{W}$, $\bar{f}(\q{W}, \boldsymbol{\mu}_S)$ is a concave differentiable function over $\boldsymbol{\mu}_S$. Thus, the optimal $\boldsymbol{\mu}_S$ is calculated by $\frac{\partial \bar{f}(\q{W}, \boldsymbol{\mu}_S)}{\partial \mu_{S,j}}=0$, i.e., optimal $\mu_{S,j}^* = \gamma_{S,j}$ for $j \in \{1, \ldots, N_R\}$ {\cite{zargari2025}}.}

{\textit{Optimizing $\q{W}$ for fixed $\boldsymbol{\mu}_S$:} For a given $\boldsymbol{\mu}_S$, the objective function $\bar{f}(\q{W}, \boldsymbol{\mu}_S)$ is simplified  by eliminating the constant terms with respect to $\q{W}$, reformulating the problem as }
\begin{subequations}
    \begin{eqnarray}\label{eqn_sensing_optimization_1_2}
\!\!\!	\!\!\!\!\!\!\!\!	\mathcal P_{1.1}:~&& \max_{\q{W}}  \quad \sum\nolimits_{j = 1}^{N_R} \frac{\hat{\mu}_{S,j} \Tr{\q{G}_{j,t} \q{W} \q{W}^{\mathrm{H}} \q{G}_{j,t}^{\mathrm{H}}}}{\bar{A}_j}, \label{eqn_P12_obj_2} \\
    \!\!\!	\!\!\!\!\!\!\!\!	\text{s.t} \quad && \frac{|\hat{\q{f}}_{k}^{\mathrm{H}}\q{W}\q{E}_k|^2}{\sum_{j=0, j\neq k}^{K} |\hat{\q{f}}_{k}^{\mathrm{H}}\q{W}\q{E}_j|^2 + 1} \ge \gamma_{C,k}^{\rm{th}},~ \forall k, \label{eqn_P1_const_1_1} \\ 
        \quad && \Tr{\q{W} \q{W}^{\mathrm{H}}} \le \rho N_T \label{eqn_P1_const_2_1},
    \end{eqnarray}
\end{subequations}
where $\hat{\mu}_{S,j} = 1 + \mu_{S,j}$, $\bar{A}_j = \Tr{\q{G}_{j,t} \q{W} \q{W}^{\mathrm{H}} \q{G}_{j,t}^{\mathrm{H}}} + \sum_{c=1}^{N_C} \Tr{\q{G}_{j,c} \q{W} \q{W}^{\mathrm{H}} \q{G}_{j,c}^{\mathrm{H}}} + M_R$, and $\gamma_{C,k}^{\rm{th}} \triangleq 2^{\mathcal{R}_{C}^{\rm{th}}/\kappa} - 1$.

{To solve $\mathcal{P}_{1.1}$, we introduce a modified matrix $\q{V} = [\q{v}_0, \q{v}_1, \cdots, \q{v}_K]$ to deal with the inequality and to normalize the constraint in {\eqref{eqn_P1_const_2_1}}, yielding $\Tr{\q{V} \q{V}^{\mathrm{H}}} = \Tr{\q{W} \q{W}^{\mathrm{H}}} + \Vert \q{z} \Vert_2^2$, where $\q{v}_k = [\q{w}_k^{\mathrm{T}}, z_k]^{\mathrm{T}}$ and $\q{z} = [z_0, z_1, \cdots, z_K]^{\mathrm{T}}$ is an auxiliary vector that simplifies the power normalization without altering the constraint {\cite{zargari2025}}. This leads to a complex sphere manifold $\mathcal{M}_S$, i.e., $\mathcal{M}_S = \{ \q{V} \in \mathbb{C}^{(N_T M_T + 1) \times (K+1)}\;| \; \Tr{\q{V} \q{V}^{\mathrm{H}}} = 1 \}$. By defining $\tilde{\q{G}}_{j,m} = \sqrt{\rho N_T}[\q{G}_{j,m}, \q{0}_{M_R \times 1}] \in \mathbb{C}^{M_R \times (N_T M_T + 1)}$ for $m \in \{t, c\}$, and $\tilde{\q{f}}_{k} = \sqrt{\rho N_T} [\hat{\q{f}}_{k}^{\mathrm{T}}, 0]^{\mathrm{T}} \in \mathbb{C}^{(N_T M_T + 1) \times 1}$ to match the dimensionality of $\q{V}$, $\mathcal{P}_{1.1}$ is reformulated as a constraint optimization problem on $\mathcal{M}_S$, i.e., }
\begin{eqnarray}\label{eqn_sensing_optimization_1_3}
    \mathcal P_{1.2}: \min_{\q{V} \in \mathcal{M}_S}  \quad \hat{f}(\q{V}), \quad \text{s.t} \quad \hat{u}_k(\q{V}) \le 0,~ \forall k,
\end{eqnarray} 
where $\hat{f}(\q{V}) = - \sum_{j = 1}^{N_R} \frac{\hat{\mu}_{S,j} \mathrm{Tr}{(\tilde{\q{G}}_{j,t} \q{V} \q{V}^{\mathrm{H}} \tilde{\q{G}}_{j,t}^{\mathrm{H}})}}{\hat{A}_{1,j}}$, $\hat{A}_{1,j} = \mathrm{Tr}{(\tilde{\q{G}}_{j,t} \q{V} \q{V}^{\mathrm{H}} \tilde{\q{G}}_{j,t}^{\mathrm{H}})} + \sum_{c=1}^{N_C} \mathrm{Tr}{(\tilde{\q{G}}_{j,c} \q{V} \q{V}^{\mathrm{H}} \tilde{\q{G}}_{j,c}^{\mathrm{H}})} + M_R$, $\hat{u}_k(\q{V}) = \gamma_{C,k}^{\rm{th}} - \frac{|\tilde{\q{f}}_{k}^{\mathrm{H}}\q{V}\q{E}_k|^2}{\hat{A}_{2,k}}$, and $\hat{A}_{2,k} \!=\! \sum_{j=0, j\neq k}^{K} |\tilde{\q{f}}_{k}^{\mathrm{H}}\q{V}\q{E}_j|^2 \!+\! 1$.

{In $\mathcal{P}_{1.2}$, $\hat{f}(\q{V})$ and $\hat{u}_k(\q{V})$ are twice continuously differentiable functions from $\mathcal{M}_S$ to $\mathbb{R}$, where $\mathcal{M}_S$ is a Riemannian manifold. Since $\mathcal{P}_{1.2}$ has an additional constraint beyond the manifold constraint, we employ the ALM to handle constraint $\hat{u}_k(\q{V}) \le 0$. Specifically, ALM constructs an augmented Lagrangian by combining the objective function $\hat{f}(\q{V})$ and constraint $\hat{u}_k(\q{V}) \le 0$, while dynamically adjusting the penalty to balance constraint enforcement and optimization robustness {\cite{liu2020b, Birgin2014book, zargari2025}}. The cost function is given as}
 \vspace{-2mm}
\begin{equation}\label{eqn_sensing_cost_function}
    \mathcal{L}_{\zeta_S}(\q{V}, \boldsymbol{\lambda}_S) = \hat{f}(\q{V}) + \frac{\zeta_S}{2} \sum_{k = 1}^{K} \max \Big\{ 0, \frac{\lambda_{S,k}}{\zeta_S} + \hat{u}_k(\q{V}) \Big\}^2,
\end{equation}
{where $\zeta_S > 0$ is a penalty parameter and $\boldsymbol{\lambda}_S  = [\lambda_{S,1}, \cdots, \lambda_{S,K}] \in \mathbb{R}^K$ is the Lagrange multiplier vector with $\lambda_{S,k} \ge 0,~\forall k$. ALM proceeds iteratively by first optimizing $\q{V}$ for a fixed $\boldsymbol{\lambda}_S$ through an unconstrained search on $\mathcal{M}_S$, followed by updating $\boldsymbol{\lambda}_S$ {\cite{Birgin2014book}}. The resulting optimization problem can be written  as}
\vspace{-2mm}
\begin{eqnarray}\label{eqn_sensing_optimization_1_4}
    \mathcal P_{1.3}:~&& \min_{\q{V} \in \mathcal{M}_S, \boldsymbol{\lambda}_S}  \quad \mathcal{L}_{\zeta_S}(\q{V}, \boldsymbol{\lambda}_S).
\end{eqnarray}
  \vspace{-5mm}

{Optimizing $\mathcal P_{1.3}$ on $\mathcal{M}_S$ involves the following steps {\cite{liu2020b}}: (i) Riemannian gradient computation, (ii) Search direction, (iii) Retraction (mapping), and (iv) Lagrange multiplier update. For details of these steps, insights, and algorithmic details, interested readers are referred to {\cite{liu2020b, zargari2025, zargari2024}}, and we omit them for brevity. Moreover, the Euclidean gradient required for the Riemannian gradient computation of the objective function {\eqref{eqn_sensing_optimization_1_4}} is given in {\eqref{eqn_sensing_const_function_grad}}.}
\begin{figure*}
    \begin{eqnarray}\label{eqn_sensing_const_function_grad}
    \!\!	\nabla_{\q{V}^{(t)}} \mathcal{L}_{\zeta_S}(\q{V}, \boldsymbol{\lambda}_S)\!\!&=&\!\! -2 \sum_{j=1}^{N_R} \hat{\mu}_{S,j} \Bigg(\frac{\tilde{\q{G}}_{j,t}^{\mathrm{H}} \tilde{\q{G}}_{j,t} \q{V}}{\hat{A}_{1,j}} - \frac{\Tr{\tilde{\q{G}}_{j,t} \q{V} \q{V}^{\mathrm{H}} \tilde{\q{G}}_{j,t}^{\mathrm{H}}}}{\hat{A}_{1,j}^2} \Big(\tilde{\q{G}}_{j,t}^{\mathrm{H}} \tilde{\q{G}}_{j,t} \q{V} + \sum_{c=1}^{N_C} \tilde{\q{G}}_{j,c}^{\mathrm{H}} \tilde{\q{G}}_{j,c} \q{V}\Big)\Bigg) \nonumber \\
        && \!\!\hspace{-8mm} - 2 \zeta_S \sum_{k=1}^{K} \q{1}_{\big\{\frac{\lambda_{S,k}}{\zeta_S} + \hat{u}_k(\q{V})\big\}} \Big(\frac{\lambda_{S,k}}{\zeta_S} + \hat{u}_k(\q{V}) \Big) \Bigg( \frac{\tilde{\q{f}}_k^{\mathrm{H}} \q{V} \q{E}_k \tilde{\q{f}}_k \q{E}_k^{\mathrm{H}}}{\hat{A}_{2,k}} - \frac{|\tilde{\q{f}}_{k}^{\mathrm{H}}\q{V}\q{E}_k|^2}{\hat{A}_{2,k}^2}\sum_{j=0, j \neq k}^{K} \tilde{\q{f}}_k^{\mathrm{H}} \q{V} \q{E}_j \tilde{\q{f}}_k \q{E}_j^{\mathrm{H}} \Bigg) \quad
    \end{eqnarray}
    
    \vspace{-2mm}
     
    \hrulefill
    \vspace{-6mm}
\end{figure*}

\vspace{-5mm}
\balance
\bibliographystyle{IEEEtran}
\bibliography{IEEEabrv,References}

\end{document}